\documentclass[11pt]{article} 

\usepackage{authblk,blindtext}
\usepackage{natbib}
\usepackage{amsmath,amssymb,amsthm,upgreek}
\usepackage{graphicx,float}
\usepackage[notref,final]{showkeys}  

\usepackage[margin=1in]{geometry}

\newtheorem{theorem}{Theorem}

\newtheorem{assum}{Assumption}
\newtheorem{defin}{Definition}
\newtheorem{remark}{Remark}
\newtheorem{propo}{Propositon}

\newtheorem{lemma}{Lemma}

\DeclareMathOperator*{\argmax}{arg\,max}

\newcommand{\eps}{\epsilon}
\newcommand{\bv}{\boldsymbol{v}}
\newcommand{\bV}{\boldsymbol{V}}

\newcommand{\ba}{\boldsymbol{a}}
\newcommand{\bb}{\boldsymbol{b}}
\newcommand{\tba}{\tilde{\boldsymbol{a}}}
\newcommand{\tbb}{\tilde{\boldsymbol{b}}}
\newcommand{\be}{\boldsymbol{e}}

\newcommand{\dd}{\,\textrm{d}}

\newcommand{\bR}{\boldsymbol{R}}

\newcommand{\bk}{{\boldsymbol{k}}}

\newcommand{\bx}{\boldsymbol{x}}

\newcommand{\ii}{\text{i}}

\newcommand{\maxsup}{^{\hh\text{\tiny max}}}
\newcommand{\minsub}{_{\text{\tiny min}}}

\newcommand{\bzero}{\boldsymbol{0}}

\newcommand{\bnu}{\boldsymbol{\nu}}

\newcommand{\psup}{^{\mbox{\tiny ($p$)}}}

\newcommand{\beq}{\begin{equation}}
\newcommand{\eeq}{\end{equation}}
\newcommand{\bali}{\begin{aligned}}
\newcommand{\eali}{\end{aligned}}

\newcommand{\hhs}{\hspace*{0.3pt}}
\newcommand{\hh}{\hspace*{0.7pt}}
\newcommand{\nes}{\hspace*{-0.2pt}}
\newcommand{\nex}{\hspace*{-0.4pt}}
\newcommand{\nesx}{\hspace*{-0.7pt}}

\newcommand{\sip}{\!\cdot\!}
\newcommand{\sir}{\nes\cdot\nes}

\newcommand{\shp}{\nes+\nes }

\newcommand{\mcS}{\mathcal{S}}
\newcommand{\mcSd}{\mathcal{S}_\delta}

\newcommand{\bkn}{\bk_n}
\newcommand{\btaun}{\tau_{\nes n}}

\usepackage{amssymb,MnSymbol,amsmath}
\usepackage{amssymb,MnSymbol,amsmath}
\usepackage{accents}

\newcommand{\bsig}{\boldsymbol{\sigma}}
\newcommand{\bupsig}{\boldsymbol{\upsigma}}

\newcommand{\shh}{\nes=\nes}

\newcommand{\bxi}{\boldsymbol{\xi}}

\newcommand{\es}{\hspace*{0.7pt}}
\newcommand{\dY}{\,\text{d}Y_{\nes\bx}}

\newcommand{\dYt}{\,\text{d}\tilde{Y}_{\nes\tilde{\bx}}}

\newcommand{\dYs}{\,\text{d}Y_{\nes\bxi}}

\graphicspath{{figures/}}

\begin{document} 

\title{On the surface Bloch waves in truncated periodic media: scalar-wave primer}

\author[1]{Bojan B. Guzina\thanks{Corresponding author: guzin001@umn.edu}}
\author[2]{Shixu Meng}
\author[1]{Prasanna Salasiya}
\author[1]{Long Nguyen}

\affil[1]{Department of Civil, Environmental, and Geo- Engineering, University of Minnesota, Minneapolis, MN, US }
\affil[2]{Mathematical Sciences, The University of Texas at Dallas, Richardson, TX 75025, U.S.}

\maketitle

\begin{abstract}
\noindent Much like their counterparts in homogeneous elastic solids, waves in periodic media can be broadly classified into Floquet-Bloch \emph{body waves}, and evanescent \emph{surface waves}. Our goal is to elucidate the latter boundary layers, termed surface Bloch (SB) waves, affiliated with rational surface cuts and homogeneous Neumann data. To this end we adopt a two-dimensional (2D) scalar wave equation with periodic coefficients (describing antiplane shear waves in phononic crystals) as a test bed and develop a unit cell-of-periodicity-based, reduced order model of the SB waves that is capable of describing their dispersion, waveforms, and ``skin depth''. The centerpiece of our analysis is a quadratic eigenvalue problem (QEP) for the effective unit cell of periodicity -- deriving from a geometric interplay between the mother Bravais lattice and orientation of the surface cut -- that seeks the complex wavenumber quantifying the evanescence away from the cut plane given (i) the excitation frequency, and (ii) wavenumber in the direction of the cut plane. In this way the sought boundary layer is obtained by superposition of the evanescent QEP eigenstates, whose relative amplitudes are obtained by imposing the homogeneous boundary condition. With the QEP eigenspectrum at hand, evaluation of an SB wave -- in terms of both dispersion characteristics and evanescent waveforms -- entails only a low-dimensional eigenvalue problem. This feature caters for rapid exploration of the effect of (periodic) surface undulations, and so enables manipulation of SB waves via optimal design of the surface cut. Our analysis also includes an account for the power flow and ``skin depth'' of surface Bloch waves, both of which are critical for the energetic relevance of boundary layers. 
\end{abstract}

\maketitle 

\section{Introduction} \label{intro}

\noindent Since the late nineteenth century~\cite{bose1898rotation}, locally-periodic structures have been devised  \citep[e.g.][]{kock1948metallic,cummer2016controlling, kadic20193d} to control ``macroscopic'' waves whose wavelength exceeds the characteristic lengthscale of medium periodicity. Commonly, such structures -- often referred to as metamaterials -- are designed through the prism of their dispersive characteristics, which concern the wave motion in an infinite periodic medium. Yet, the metamaterials' capacity as wave regulators is inevitably brought to bare via interaction of their \emph{bounded subsets} with the environment \cite{assouar2018acoustic,zangeneh2020topological,zhu2013acoustic}.  

Historically, the analyses of wave motion in unbounded periodic media have been pursued via the Floquet-Bloch expansion \cite[e.g.][]{bensoussan2011asymptotic, brillouin1946wave, kuchment2012floquet, wilcox1978theory}. This approach results in a dispersion map relating the frequency and wavenumber of a Floquet-Bloch wave, commonly obtained by solving a linear eigenvalue problem~\cite{bensoussan2011asymptotic} over the unit cell of medium periodicity. In recent years, Floquet-Bloch waves have been used to develop reduced-order models for a variety of wave scattering problems \cite[e.g.][]{hussein2009reduced, krattiger2018generalized, tsukamoto2014real} and help establish the convergence results of relevant numerical algorithms \cite{fliss2016solutions, lechleiter2017floquet, kirsch2023periodic}. Novel applications of metamaterials, however, necessitate the computation of an extended dispersion relationship that accounts for complex wave vectors, e.g. those inside a band gap. This can be achieved by prescribing the frequency of a Floquet-Bloch wave and seeking the admissible wave vector component(s) -- which results in a quadratic eigenvalue problem (QEP) \cite{tassilly1987propagation,tisseur2001quadratic,hussein2014dynamics} over the unit cell of periodicity. Recently, such an approach to computing the complex dispersion relationship has attracted much attention in mathematics \cite{engstrom2009spectrum, Lack2019, palermo2020reduced}, physics \cite{brule2016calculation,raman2010photonic,rybin2016inverse}, and engineering  \cite{hussein2014dynamics, hussein2009theory, toader2004photonic}. When considering the wave motion in bounded subsets of periodic media, on the other hand, one must account for the emergence of \emph{boundary layers} whose analysis is considerably more difficult \cite[e.g.][]{cakoni2019scattering, joly2006exact, Kulpe2015,doi:10.1137/17M1118920}. Typically, the evanescent (surface or interface) waves arising in this class of problems are evaluated implicitly via suitable computational scheme~\cite{fliss2010computation} and deemed as ``waves to avoid''~\cite{doi:10.1137/17M1118920}. Even in the works where the boundary layers are addressed explicitly, their analysis is either (i) not simpler than that of the full boundary value problem~\cite{cakoni2016homog, cakoni2019scattering}, or (ii) based on numerical, supercell-driven evaluation~\cite[e.g.][]{charara2023cell,fang2024coupled,wang2015topological}  which sheds little light on the problem. By allowing for complex wave vectors, however, the ``quadratic'' Bloch eigenstates solving the QEP have recently emerged as viable building block supporting the reduced-order models of boundary layers in periodic media~\cite{Kulpe2015,willis2016negative,salasiya2024plug}. 

To date, there are no comprehensive analytical and computational frameworks elucidating the boundary layers -- and in particular surface waves -- in periodic media. Even within the realm of topological mechanics~\cite{huber2016topological,xin2020topological} (governed by differential equations), the analysis of interfacial states revolves around the topological invariants introduced by analogy with topological physics~\cite{bernevig2013topological} (governed by Hamiltonians). For instance, topological invariants may reveal which (evanescent) interfacial states may be topologically protected from scattering-induced losses, but they furnish no information about the ``skin depth'' of such boundary layers -- that is critical for meaningful energy tunneling. Similarly, tools are available to manipulate the surface waves in homogeneous media by grating or periodically decorating (with resonators) their free surfaces~\cite[e.g.][]{colombi2016seismic, colquitt2017seismic}; yet, little is known about the possibility of manipulating boundary layers propagating along the free surface of a periodic medium. 

In this vein our goal is to understand, simulate, and control the  boundary layers propagating along the free surface of a periodic medium, characterized by the homogeneous Neumann boundary condition. For the sought surface wave to retain its (periodic) Bloch character, we require the surface cut to make a rational slope relative to the Bravais lattice of a periodic medium.  As a test bed for the analysis, we consider the scalar wave equation with periodic coefficients. In the context of multi-dimensional problems, the very ability of the QEP to produce complex wave vectors allows one to identify and compute the evanescent Bloch eigenstates that propagate (resp. decay) in the direction parallel (resp. orthogonal) to the boundary. This motivates us to pursue the sought boundary layer, termed surface Bloch wave, as a linear combination of suitably-defined QEP eigenstates satisfying the homogeneous Neumann boundary condition.  With the ``mother'' QEP eigenspectrum at hand, evaluation of the surface Bloch wave -- in terms of both dispersion characteristics and evanescent waveforms -- entails only a low-dimensional eigenvalue problem. This in turn caters for a rapid exploration of the design space in terms of (periodic) undulations of the free surface, and so opens the door toward controlling boundary layers in periodic media via optimal design of the surface cut. For completeness, our analysis includes an account for the power flow and ``skin depth'' of a surface Bloch wave, both of which are critical for the energetic relevance of boundary layers. 

For completeness, we remark that a similar term, ``Bloch surface waves'', has appeared earlier~\cite[e.g.][]{liscidini2007enhancement} in the context of photonics involving one-dimensional (1D) periodic arrangements of homogeneous strata. Another akin expression, ``Rayleigh-Bloch waves'', refers to the boundary layers due to diffraction gratings~\cite[e.g.][]{wilcox2012scattering} generated by either (i) grating the boundary of, or (ii) inserting a strip of a 1D-periodic lattice in, an otherwise homogeneous medium~\cite[e.g.][]{antonakakis2014asymp,colquitt2015rayleigh, chaplain2025acoustic}.  

\section{Preliminaries} \label{prelim}

\noindent Consider a 2D periodic medium supported on Bravais lattice $\bR\subset\mathbb{R}^2$ featuring the primitive (covariant) vector basis $\ba_i$ ($i\!=\!\overline{1,2}$). To facilitate the analysis, we also  (i) introduce the contravariant vector basis $\bb^{j}$ ($j\nes\in\nes\overline{1,2}$) satisfying $\ba_i \cdot \bb^j \!= \delta_{ij}$ where $\delta_{ij}$~is the Kronecker delta, and (ii) assume Einstein summation notation over repeated indices $i,j\!=\!\overline{1,2}$ (unless stated otherwise). On denoting by \mbox{$x^{j}=\bx\nes\cdot\nes\bb^j$}  the contravariant components of the position vector $\bx\in\mathbb{R}^2$, the featured Bravais lattice and unit cell of periodicity~$Y$  can be written respectively as 
\begin{equation}\label{brav}
\bR \;=\; \big\{m^{\nes i} \ba_i:\; m^i\!\in\mathbb{Z}\big\}, \qquad 
Y =\, \big\{\bx = x^i\hh\ba_i:\: -\tfrac{1}{2}< x^i\!<\tfrac{1}{2}\big\}.  
\end{equation}  
The reciprocals of~$\bR$  and~$Y$, which characterize the wavenumber space, are given respectively by   
\begin{equation}\label{brav*}
\bR^* \!=\; \big\{2\pi (m_{j} \bb^j):\; m_{j}\!\in\mathbb{Z}\big\}, \qquad 
B \,:=\, Y^* \!=\, \big\{\bk = k_j\,\bb^j :\: -\pi< k_j\nes<\pi \big\}, 
\end{equation}
where $\bk$ is the wave vector, $k_j = \bk\sir\ba_j$, and~$B$ denotes the first Brillouin zone. For future reference, we also let $\be_j$ ($j\in\overline{1,2}$) denote the orthonormal Cartesian basis of $\mathbb{R}^2$.

\subsection{Bloch waves in an unbounded periodic medium}\label{trans}

\noindent Our focus is on the time-harmonic scalar wave equation with periodic coefficients, namely 
\begin{equation}\label{PDE1}
\nabla \sip (G(\bx) \nabla u) + \omega^2\rho(\bx)\hh u \:=\: 0, \qquad \bx \in \mathbb{R}^2
\end{equation}
where $G$ and $\rho$ are $Y$-periodic; $\omega$ is the frequency of oscillations, and the wavefield $u(\bx)$ carries implicit time dependence~$e^{-\ii\omega t}$. Hereon, we assume that $G$ and~$\rho$ are real-valued $L^\infty(Y)$ functions bounded from below away from zero in that $G_{\inf} \leqslant G \leqslant G_{\sup}$ and $\rho_{\inf} \leqslant \rho \leqslant \rho_{\sup}$ for some positive constants $G_{\inf}, G_{\sup}, \rho_{\inf}$ and~$\rho_{\sup}$. In physical terms \eqref{PDE1} can be interpreted as the model for antiplane shear (SH) wave motion in phononic crystals, where  $G,\rho$ and $u$ signify respectively the shear modulus, mass density, and out-of-plane transverse displacement. Implicitly, this model assumes small-amplitude vibrations and no internal dissipation, resulting in a linear field equation with real-valued elastic modulus~$G$. 

Recalling the Floquet-Bloch theorem \citep{Floq1883,Bloch1929}, we factor the wavefield~$u$ as 
\begin{equation} \label{floquet}
u(\bx) \:=\: \phi(\bx) \, e^{\ii \bk \cdot \bx} \:=\: \phi(\bx) \, e^{\ii k_j x^j}, \qquad \phi:\: \text{$Y$-periodic}
\end{equation} 
where $\bk\in\mathbb{C}^2$ is the wave vector, and $\phi$ depends implicitly on~$\omega$ and~$\bk$. From~\eqref{PDE1} and~\eqref{floquet}, we obtain 
\begin{equation} \label{EVP}
\begin{aligned}
& \text{(a) ~~} \qquad \nabla_{\!\bk}  \sip (G(\bx) \nabla_{\!\bk} \phi) + \omega^2 \rho(\bx) \phi = 0, \quad \bx \in Y  \\*[1mm]
& \text{(b) ~~} \phi|_{x^i=-\frac{1}{2}} = \phi|_{x^i=\frac{1}{2}}, \quad~
\bnu \sip G\nabla_{\!\bk}\phi|_{x^i=-\frac{1}{2}} = -\bnu \sip G\nabla_{\!\bk}\phi|_{x^i=\frac{1}{2}}, \quad~ i\shh \overline{1,2} 
\end{aligned}
\end{equation}
where $\bnu = \nu^i \ba_i$ is the unit outward normal on~$\partial{Y}$ and 
\begin{equation} \label{bk}
\nabla_{\!\bk}  \hh:=\, \nabla \!+ \ii \bk \,=\, \Big(\frac{\partial}{\partial x^j} + \ii k_j\Big) \bb^j. \end{equation}

\subsection{Quadratic eigenvalue problem (QEP)} \label{secQEP}

\noindent To help formulate the problem, we introduce the Hilbert spaces
\begin{eqnarray*}
L_p^2(Y)  &\!\shh \!&  \{g\in L^2(Y):\, (g, g) < \infty, \: g|_{x^i=-\frac{1}{2}} = g|_{x^i=\frac{1}{2}} \}, \\
H^1_{p}(Y) &\! \shh \!& \{g\in L^2(Y):\, \nabla g \in (L^2(Y))^2, \: g|_{x^i=-\frac{1}{2}} = g|_{x^i=\frac{1}{2}}\}
\end{eqnarray*}
where~$(g,h)=\int_{Y}g(\bxi)\es\overline{h}(\bxi)\dYs$ and $\overline{h}$ is the complex conjugate of~$h$. By virtue of a test function $\psi\!\in\! H^1_{p}(Y)$, a weak statement of~(\ref{EVP}a) can be computed as 
\begin{equation} \label{EVP-weak}
\int_Y G\, \nabla_{\!\bk} \phi \es\cdot\nes \nabla_{\!\nes-\bk} \overline{\psi}\, \dY \:-\: \omega^2\!\! \int_Y \rho\, \phi \, \overline{\psi}\, \dY \:=\: 0 \qquad \forall\psi \in H_p^1(Y),
\end{equation}
subject to the boundary conditions~(\ref{EVP}b).

On recalling the wave vector decomposition $\bk=k_j\bb^j$ and introducing indices $\alpha\nes\in\{1,2\}$ and $\beta\shh\{1,2\}\,\backslash\hh\{\alpha\}$, we next pose an eigenproblem that consists in seeking the eigenvalues $\kappa \nes:=\nes k_\alpha \in\mathbb{C}$ and eigenfunctions $\phi\in H_p^1(Y)$ that satisfy~\eqref{EVP-weak} for given $\omega$ and~$k_\beta$. In this setting, we obtain the skew-lattice QEP 
\begin{multline}  \label{QEP-surf1}
\kappa_n^2  \hh\Gamma^{\alpha\alpha} \!\! \int_{Y} G \hh \phi_n\es \overline{\psi} \hh\dY \,+\, 
\kappa_n \bigg\{2 k_\beta \hh \Gamma^{\alpha\beta}\!\! \int_{Y} G\hh\phi_n\,\overline{\psi} \hh\dY \hh-\hh \ii\,\Gamma^{\alpha j}\!\! \int_{Y} G \left( \frac{\partial \phi_n}{\partial x^j} \overline{\psi} - \phi_n \frac{\partial \overline{\psi}}{\partial x^j} \right)\dY \bigg\} \\ \,+\, 
\bigg\{\Gamma^{\beta\beta} k_{\beta}^2 \!\int_{Y} G\hh \phi_n\es\overline{\psi} \hh\dY
\,-\, \ii  k_\beta \Gamma^{\beta j} \!\!
\int_{Y} G \left( \frac{\partial \phi_n}{\partial x^j} \overline{\psi} - \phi_n \frac{\partial \overline{\psi}}{\partial x^j} \right)\dY \\
+ \int_{Y} \Big(G\hh \nabla \phi_n \nes\cdot\nes \nabla \overline{\psi} - \omega^2\! \rho\hh\phi_n\es \overline{\psi}\Big) \hh\dY \bigg\}  \:=\: 0 \qquad \forall\psi \in H_p^1(Y),
\end{multline}
where $\Gamma^{ij}\!=\bb^i\!\cdot\nes\bb^j$ are the components of the (symmetric, positive definite) contravariant metric tensor, and implicit summation is assumed over the repeated index $\,j=\overline{1,2}$.

\begin{remark}
 Since the ``input/output'' indices~$\beta$ and~$\alpha$ are fixed from the onset, our implicit summation convention over repeated indices applies to symbols $i,j=\overline{1,2\,}$ but not $\alpha,\beta\in\{1,2\}$.
\end{remark}

For an orthogonal lattice in~$\mathbb{R}^2$ with $\beta=1$, referring to the Cartesian coordinates~$\xi_j$ we have $x^j\mapsto \xi_j,\Gamma^{ij}\mapsto \delta_{ij}$, and $\alpha=2$ which reduces~\eqref{QEP-surf1} to an earlier result~\cite{salasiya2024plug}, namely 
\begin{multline}  \label{QEP-surf2}
\kappa_n^2 \int_{Y} G \hh \phi_n\es \overline{\psi} \hh\dYs \,-\, 
\kappa_n \left\{\ii\!\!\int_{Y} G \left( \frac{\partial \phi_n}{\partial \xi_2} \overline{\psi} - \phi_n \frac{\partial \overline{\psi}}{\partial \xi_2} \right)\dYs \right\} \,+\, 
\bigg\{k_1^2 \!\int_{Y} G\hh \phi_n\es\overline{\psi} \hh\dYs \\ 
- \ii \hh k_1 \! \int_{Y} G \left( \frac{\partial \phi_n}{\partial \xi_1} \overline{\psi} - \phi_n \frac{\partial \overline{\psi}}{\partial \xi_1} \right)\dYs 
+ \int_{Y}\Big(G\hh \nabla \phi_n \nes\cdot\nes \nabla \overline{\psi} \,-\, \omega^2\!\rho\hh\phi_n\es \overline{\psi}\Big)\hh\dYs  \bigg\}  \:=\: 0 \qquad \forall\psi \in H_p^1(Y).  
\end{multline} 

\begin{remark}
In this study, we seek the boundary layers in periodic media as evanescent Bloch waves that decay exponentially away from the boundary. This motivates the introduction of~\eqref{QEP-surf1} where for a given ``propagating" wave vector component $k_\beta\in\mathbb{R}$, the sought QEP eigenvalue $\kappa\shh k_\alpha\in\mathbb{C}$ is precisely the component of~$\bk$ that controls the evanescence of~$u(\bx)$ via~\eqref{floquet}. This claim will be elaborated on in Section~\ref{SBW}\ref{SBW-QEP}.
\end{remark}

\subsection{Analysis of the QEP} \label{QEProps}

\noindent Letting $\langle \cdot,\cdot \rangle$ denote the inner product affiliated with $H^1_{p} (Y)$, we introduce operators
$\mathcal{A}:H^1_{p} (Y)  \mapsto H^1_{p} (Y) $, $\, \mathcal{B}:H^1_{p}
(Y)  \mapsto H^1_{p} (Y) $ and $\, \mathcal{C}:H^1_{p} (Y)  \mapsto
H^1_{p} (Y)$ respectively by  
\begin{eqnarray*}
\langle \mathcal{A} \phi, \psi \rangle \!\!\!&:=&\!\!\! \Gamma^{\beta\beta} k_{\beta}^2 \!\int_{Y} G\hh \phi \es\overline{\psi} \hh\dY
- \ii  k_\beta \Gamma^{\beta j} \!\!
\int_{Y} G \left( \frac{\partial \phi}{\partial x^j} \overline{\psi} - \phi \frac{\partial \overline{\psi}}{\partial x^j} \right)\dY   \nonumber \\
&&+ \int_{Y} \big(G\hh \nabla \phi \nes\cdot\nes \nabla \overline{\psi} - \omega^2\! \rho\hh\phi\hh \overline{\psi}\Big) \hh  \dY, \quad \forall \phi,\psi \in H^1_{p} (Y),\\
\langle \mathcal{B} \phi, \psi \rangle \!\!\!&:=&\!\!\! 2 k_\beta \hh \Gamma^{\alpha\beta}\!\! \int_{Y} G\hh\phi\,\overline{\psi} \hh\dY \hh-\hh \ii\,\Gamma^{\alpha j}\!\! \int_{Y} G \left( \frac{\partial \phi}{\partial x^j} \overline{\psi} - \phi \frac{\partial \overline{\psi}}{\partial x^j} \right)\dY,\quad \forall \phi,\psi \in H^1_{p} (Y),\\
\langle \mathcal{C} \phi, \psi \rangle \!\!\!&:=&\!\!\!  \Gamma^{\alpha\alpha} \!\! \int_{Y} G \hh \phi\hh \overline{\psi} \hh\dY ,\quad \forall \phi,\psi \in H^1_{p} (Y).
\end{eqnarray*}
With such definitions, QEP \eqref{QEP-surf1} becomes  
\begin{equation}\label{QEP-lam}
\mathcal{A} \phi_n + \kappa_n\hh \mathcal{B} \phi_n + \kappa_n^2\hh \mathcal{C} \phi_n \,=\, 0.
\end{equation}
To aid the ensuing developments, we establish several key properties of~\eqref{QEP-lam} via the following claims whose proofs are provided in Appendix~A, electronic supplementary material (ESM).

\begin{lemma} \label{lemma1}
Operators $\mathcal{A}:H^1_{p} (Y)  \mapsto H^1_{p} (Y) $, $\, \mathcal{B}:H^1_{p} (Y)  \mapsto H^1_{p} (Y) $ and $\,\mathcal{C}:H^1_{p} (Y)  \mapsto H^1_{p} (Y) $ are self-adjoint.
\end{lemma}

\begin{lemma}  \label{lemma2}
Operator $\mathcal{B}:H^1_{p} (Y)  \mapsto H^1_{p} (Y) $ is compact, and operator 
\[
\mathcal{T}(\lambda) = \mathcal{A}   + \lambda \mathcal{B} + \lambda^2 \mathcal{C}  
\] 
is Fredholm of index zero.
\end{lemma}

\begin{propo} \label{prop1}
If $\mathcal{T}(\lambda) \hh\phi=0\, $ for some non-zero $\phi \in H^1_p(Y)$, there must exist a nontrivial $\psi \in H^1_p(Y)$ such that $\mathcal{T}(\overline{\lambda})\hh \psi=0$. Hence all eigenvalues of QEP~\eqref{QEP-surf1} come in complex-conjugate pairs.
\end{propo}

\begin{theorem}  \label{thm1}
Assume that there exists $\tau \in \mathbb{C}$ such that $\mathcal{T}(\tau)$ is injective. Then the eigenvalues $\lambda$ affiliated with the quadratic pencil $\mathcal{T}(\lambda)$  form a discrete set, with infinity being the only possible accumulation point.
\end{theorem}

\section{Surface Bloch (SB) waves} \label{SBW}

 \noindent We next consider the propagation of surface waves in a periodic medium that is truncated along a cut plane $\mcS=\mathbb{R}$ endowed with unit outward normal~$\bnu_{\nesx\mcS}$. Across the cut plane we assume the traction-free, i.e.~homogeneous Neumann, boundary condition 
\begin{equation} \label{homNeumann}
\bnu_{\nesx\mcS} \cdot \big(G(\bx) \nabla u\big)|_{\mcS}  \,=\, 0.   
\end{equation}
In this setting, we seek to expose the Bloch wave-like boundary layer(s) in a semi-infinite domain $\Omega_{\mcS}\!\subset\mathbb{R}^2$ occupying the $-\bnu_{\nesx\mcS}$ side of~$\mcS$. For specificity, the surface wave is taken to propagate in a direction given by the unit vector $\be\nes\perp\nes \bnu_{\nesx\mcS}$. For such boundary layer to have a Bloch-like character, however, $\be$ and~$\mcS$ must satisfy the following condition.

\begin{figure}[ht] 
\centering
\includegraphics[width=0.67\linewidth]{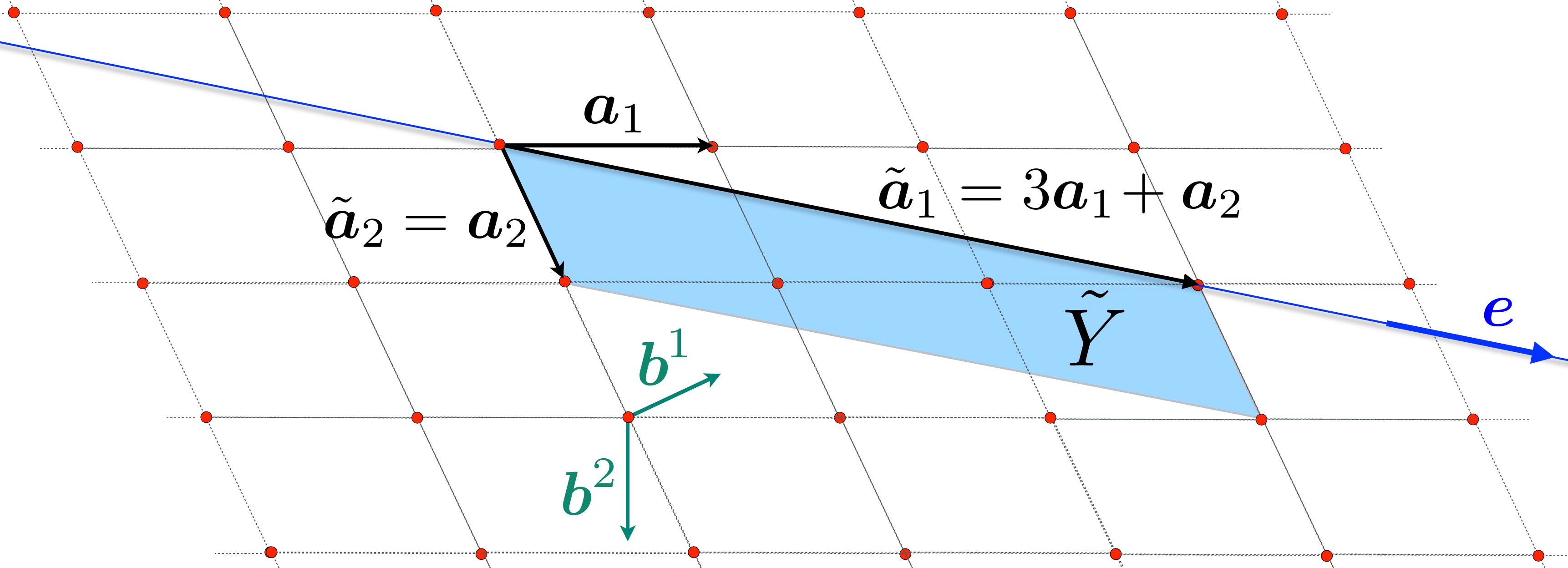}
\caption{Direction of propagation, $\boldsymbol{e}\in\mathbb{R}^2$, of a surface Bloch wave featuring a ``1:3'' rational slope ($q^1\!=3,\,q^2\!=1$) relative to lattice~$\bR$. The new lattice basis and unit cell catering for the surface wave analysis are~$(\tba_1,\tba_2)$ and~$\tilde{Y}$, respectively.} \label{slope} 
\end{figure}

\begin{defin}\label{def1}
The cut surface~$\mcS$ and direction of propagation~$\be$ of a surface wave have ``rational slope'' relative to lattice~$\bR\hh$ if there exist $q^i\!\in\mathbb{Z}$ ($i\!=\!\overline{1,2}$) such that $\be\nes\parallel\nes q^i \ba_i$. In the sequel, we refer to~$\{q^i\}$ as the pair containing the smallest such values of~$q^i$, see Fig.~\ref{slope} for an example with~$q^1\!=3$ and~$q^2\!=1$. 
\end{defin}



\begin{assum}\label{asu1}
Hereon, our focus is on the cut planes and surface wave directions with a rational slope relative to lattice~$\bR$. An extension of the analysis to irrational slopes, that is beyond the scope of this work, is possible via the lifting (i.e. cut-and-projection) approach \cite{kozlov1979ave,gerard2012homog}. 
\end{assum}

\subsection{QEP basis of a ``rational'' surface wave} \label{SBW-QEP}

\noindent With the foregoing results in place, we assume that  
\begin{equation}
\be \parallel q^j\!\ba_j ~~\text{for some}~ \{q^j\}\!\subset\mathbb{Z}. 
\end{equation}
Recalling Section~\ref{prelim}\ref{secQEP}, we proceed by adopting a new lattice basis $\{\tba_i\}\shh\{\tba_\alpha,\tba_\beta\}$ such that 
\begin{equation}
\tba_{\beta} = q^j \!\ba_j, \qquad \tba_{\alpha} = \ba_\ell, 
\end{equation}
where  
\[
\ell \,=\, \argmax_j \frac{\bnu_{\nesx\mcS}\cdot\ba_j}{\|\ba_j\|} \quad \text{(no summation over~$j$)}
\]
is selected from the original lattice basis $\{\ba_j\}$ as to minimize the angle between $\tba_\alpha$ and~$\bnu_{\nesx\mcS}$. For clarity, we emphasize that the direction $\tba_{\beta}\parallel \be$ specifies the cut plane, whereas $\tba_{\alpha}$  is the ``depth'' direction. 

\begin{remark}
By the fact that $\tba_\beta$ is coplanar with~$\mcS$, one has $\tbb^\alpha\!\!\perp\nes\mcS$. Since  $\bk\cdot\bx \hh=\hh \tilde{k}_\alpha\hh \tilde{x}^\alpha \!+\tilde{k}_\beta\hh \tilde{x}^\beta$ and \mbox{$\tilde{x}^\alpha =\bx\cdot\tbb^\alpha$}, on the other hand, it is clear that $\tilde{k}_\alpha\!\in\mathbb{C}$ inherently quantifies the decay of evanescent Bloch modes in the direction orthogonal to~$\mcS$. In the context of QEP~\eqref{QEP-surf1}, this result elucidates our claim that the input wavenumber component $k_\beta=\bk\sip\ba_\beta$ is that in the direction of the surface cut, while the output component $\kappa\shh k_\alpha$ (the eigenvalue) controls the evanescence of the wavefield away from the cut plane.
\end{remark}

With reference to Fig.~\ref{slope}, the foregoing setup allows us to consider the unit ``multi-cell'' and companion Brillouin zone
\begin{equation} \label{multiY}
\begin{aligned}
\tilde{Y} =\, \big\{\bx = \tilde{x}^i\hh\tba_i :\: -\tfrac{1}{2}< \tilde{x}^i\!<\tfrac{1}{2}\big\}, \qquad 
\tilde{B} =\, \big\{\bk = \tilde{k}_j\,\tbb^j :\: -\pi<\nes \tilde{k}_j\nes<\pi \big\}, 
\end{aligned}    
\end{equation}
and to pursue the skew-lattice QEP~\eqref{QEP-surf1} written over $\tilde{Y}$ which seeks $\tilde{\kappa}_n\nes:=\tilde{k}_\alpha\!\in\mathbb{C}$ and~$\tilde{\phi}_n\!\in H_p^1(\tilde{Y})$ for given~$\omega$ and $\tilde{k}_\beta$. Specifically, we consider 
\begin{multline}  \label{QEP-surf3}
\tilde{\kappa}_n^2  \hh\tilde{\Gamma}^{\alpha\alpha} \!\! \int_{\tilde{Y}} G \hh \tilde{\phi}_n\es \overline{\psi} \hh\dYt \,+\, \tilde{\kappa}_n \bigg\{2 \tilde{k}_\beta \hh \tilde{\Gamma}^{\alpha\beta}\!\! \int_{\tilde{Y}} G\hh\tilde{\phi}_n\,\overline{\psi} \hh\dYt \hh-\hh \ii\,\tilde{\Gamma}^{\alpha j}\!\! \int_{\tilde{Y}} G \left( \frac{\partial \tilde{\phi}_n}{\partial \tilde{x}^j} \overline{\psi} - \tilde{\phi}_n \frac{\partial \overline{\psi}}{\partial \tilde{x}^j} \right)\dYt \bigg\} \\
\,+\, \bigg\{\tilde{\Gamma}^{\beta\beta} \tilde{k}_{\beta}^2 \!\int_{\tilde{Y}} G\hh \tilde{\phi}_n\es\overline{\psi} \hh\dYt
\,-\, \ii  \tilde{k}_\beta \tilde{\Gamma}^{\beta j} \!\!
\int_{\tilde{Y}} G \left( \frac{\partial \tilde{\phi}_n}{\partial \tilde{x}^j} \overline{\psi} - \tilde{\phi}_n \frac{\partial \overline{\psi}}{\partial \tilde{x}^j} \right)\dYt \\
+ \int_{\tilde{Y}} \big( G\hh \nabla \tilde{\phi}_n \nes\cdot\nes \nabla \overline{\psi} - \omega^2\! \rho\hh\tilde{\phi}_n\es \overline{\psi} \big) \hh\dYt  \bigg\}  \:=\: 0 \qquad \forall\psi \in H_p^1(\tilde{Y}),
\end{multline}
where $\hh\tilde{k}_\beta \shh k_e$; $\, k_e\nes\in\mathbb{R}$ is the prescribed component of the wave vector $\bk$ in direction $\be=\tba_{\beta}/\|\tba_{\beta}\|$; $\tilde{\Gamma}^{ij}\!=\tbb^i\!\cdot\nes\tbb^j$ are the components of the contravariant metric tensor, and implicit summation is assumed over the repeated index $j=\overline{1,2}$. In the sequel, we endow~$\tilde{\phi}_n$ with a unit $L_p^2(\tilde{Y})$-norm. 

\begin{remark} \label{rem2}
Assuming without loss of generality that the outward normal to the half-space is oriented s.th. $\bnu_{\nesx\mcS}\nes\cdot\tbb^\alpha\!<0$, we shall hereon slightly abuse the notation and refer to  $\{\tilde{\kappa}_{n},\tilde{\phi}_{n}\}_{n=1}^\infty$ as the subset of the full eigenspectrum containing the eigenvalues with strictly positive imaginary parts, i.e.  
\begin{equation}\label{decay}
\Im(\tilde{\kappa}_n) \hh>\hh 0,    
\end{equation}
which by~\eqref{floquet} ensures the exponential decay, $e^{-\Im(\tilde{\kappa}_n)\tilde{x}^\alpha}$, of an evanescent wave with distance from $\mathcal{S}$ into the half-space. Hereon, we arrange~$\tilde{\kappa}_n$ in the order of ascending imaginary part in that $\Im(\tilde{\kappa}_{n+1})\geqslant\Im(\tilde{\kappa}_n)$. 
\end{remark}

In this setting, our focus is on the surface Bloch wavefield propagating along~$\mcS$ with prescribed wavenumber~$k_e$, namely 
\begin{equation}\label{evan1}
u(\bx) = \sum_{n=1}^\infty \eta_n\hh \tilde{\phi}_n(\bx) e^{\ii \bkn\cdot\hh\bx} \,=\; \Big( \sum_{n=1}^\infty \eta_n\hh \tilde{\phi}_n(\bx) \, e^{\ii\es \tilde{\kappa}_n \tilde{x}^\alpha} \Big) \, e^{\ii \nes \tilde{k}_\beta \tilde{x}^\beta}, \qquad \tilde{k}_\beta = k_e
\end{equation}
where~$\eta_n\!\in\nes\mathbb{C}$ are the modal amplitudes and the  modulation term $e^{\ii\es\bk\cdot\bx}$ is conveniently parsed to separate the direction of evanescence, $\tilde{x}^\alpha$, from the surface cut direction(s) $\tilde{x}^\beta$.

\subsection{Undulated surface cut}

\noindent For brevity of notation, we the denote the \emph{in-plane coordinate} by   
\begin{equation}\label{bzeta}
\zeta \:=\: \tilde{x}^\beta 
\end{equation}
and consider a cut plane located at ``depth'' $\mathfrak{d}\in[-\tfrac{1}{2},\tfrac{1}{2})$ within~$\tilde{Y}$, namely      
\begin{equation} \label{s0}
\mcS \;=\; \big\{\bx\nes\hh:\: \bx =  \zeta \tba_\beta  +\mathfrak{d}\hh\tba_\alpha\big\}.  
\end{equation}
With such definitions, the reference unit cell $\tilde{\Upsilon}_{\!\mcS}\subset{\mathbb{R}}$ of the cut plane, which specifies the ``in-plane'' periodicity of a surface Bloch wave, can be written as 
\begin{equation}\label{s1}
\tilde{\Upsilon}_{\!\mcS} \;:=~ \mcS\cap\tilde{Y} \;=\; \big\{\zeta: \,-\tfrac{1}{2}< \zeta\!<\tfrac{1}{2}\big\}.
\end{equation}

For generality, we endow the cut surface~\eqref{s0} with an $\tilde{\Upsilon}_{\!\mcS}$-periodic Lipshitz undulation. Assuming the undulation~$\delta\mcS$ to be given by a periodic bijective map $f\!:\zeta\mapsto\mathbb{R}$, this allows us to consider surface waves propagating in direction~$\be$ along $\mcSd \hh:=\hh \mcS+\delta\mcS\hh$ subject to the boundary condition 
\begin{equation}\label{s3}
\bnu_{\!\mcSd} \cdot \big(G(\bx) \nabla u\big)|_{\mcSd}  \,=\, 0, 
 \end{equation}
where 
\begin{equation}\label{s7o}
\mcS_\delta \;=\; \big\{\bx\nes:\: \bx = \zeta \tba_\beta +(\mathfrak{d} \nes+\nes f(\zeta))\hh\tba_\alpha, \; f\nes: \tilde{\Upsilon}_{\!\mcS}\textrm{-periodic}\big\};   
\end{equation}
$\zeta$ is given by~\eqref{bzeta}, and $\bnu_{\!\mcSd}$ is the unit outward normal to~$\mcSd$. 

\begin{figure}[h!] 
\centering
\includegraphics[width=0.85\linewidth]{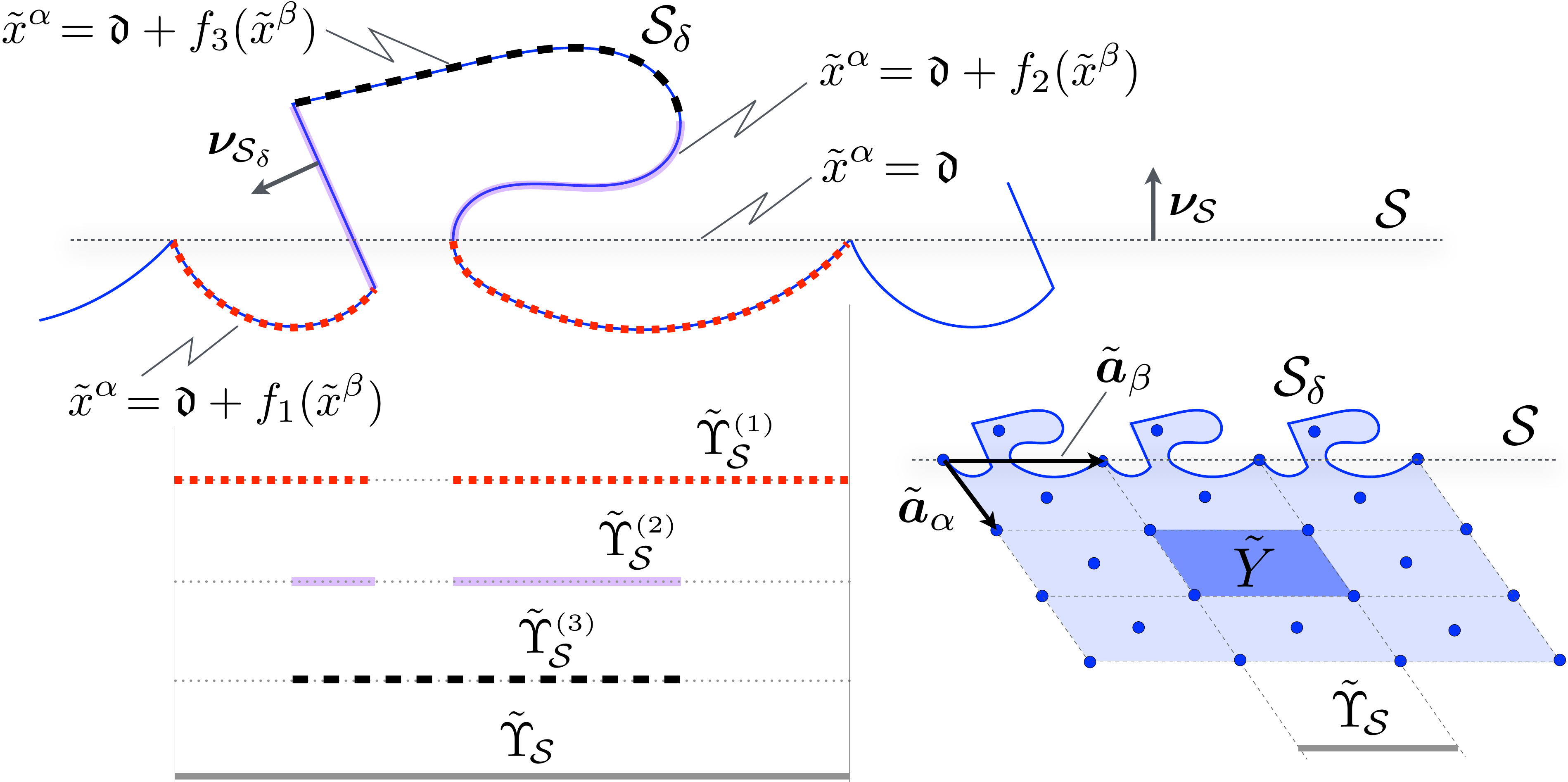}
\caption{Schematics of an undulated cut ($P=3$) at depth $\mathfrak{d}\!=\!-\tfrac{1}{2}$  with relevant features and notations. In the example, the surface cut~$\mathcal{S}$ is made at ``1:2'' rational slope ($q^1\!=2,\hh q^2\!=1$) relative to an orthogonal lattice~$\bR$. In the context of~\eqref{evan1} and~\eqref{s3}, the surface traction on the ligaments extending ``above'' the cut plane~$\mcS$ is computed via the periodicity of quadratic eigenfunctions~$\tilde{\phi}_n$. } \label{fold} 
\end{figure}

To cater for situations where $\delta\mathcal{S}$ may feature folding, one may conveniently generalize~\eqref{s7o} by describing $\mcS_\delta$ as a set of $\tilde{\Upsilon}_{\!\mcS}$-periodic bijective mappings, $f_p\nes:\tilde{\Upsilon}_{\!\mcS}\hh\supset\hh \tilde{\Upsilon}_{\!\mcS}\psup\mapsto\mathbb{R}$ as 
\begin{equation}\label{s7}
\mcS_\delta \;=\; \big\{\bx\nes:\: \bx = \zeta \tba_\beta +(\mathfrak{d} + f_p(\zeta))\hh\tba_\alpha, ~ f_p\nes: \tilde{\Upsilon}_{\!\mcS}\textrm{-periodic},~ p=\overline{1,P}\big\}, 
\end{equation}
see Fig.~\ref{fold} for an illustration with $P\shh3$.
By the finiteness of~$P$, this definition ensures that the set of points on~$\mcS_\delta$ where $|\partial \tilde{x}^\alpha/\partial\zeta|^2$ is unbounded is of zero measure in $\mathbb{R}$. This very set, together with the subset of~$\mcS_\delta$ where $\bnu_{\!\mcSd}$ is not uniquely defined (namely ``kinks'' in the cut surface) delineates the respective supports of~$\tilde{\Upsilon}_{\!\mcS}\psup$ within $\tilde{\Upsilon}_{\!\mcS}$ as shown in the figure. 

\subsection{Surface-wave eigenproblem} \label{SWEP}

\noindent By virtue of~\eqref{evan1}, the traction-free condition \eqref{s3} becomes 
\begin{equation}\label{evan2}
\sum_{n=1}^\infty \eta_n\hh \big(G(\bx)\,\bnu\cdot\nes\nabla_{\!\bkn} \nes \tilde{\phi}_n \: e^{\ii\es\bkn\cdot\hh\bx}\big)\Big|_{\mcSd} =\: 0, \qquad 
\bkn \hh=\hh \tilde{\kappa}_n\hh \tbb^{\alpha}  \!+\,  \tilde{k}_\beta\hh\tbb^{\beta}
\end{equation}
where $\bx = \tilde{x}^j\es\tba_j$, $\,\bnu_{\!\mcSd} \!= \tilde{\nu}_{\!\mcSd}^j\tba_j$, and $\nabla_{\!\bk} = (\partial/\partial \tilde{x}^j \!+\ii \es 
\tilde{k}_j) \tbb^{j}$. Recalling Remark~\ref{rem2}, we conveniently consider an approximation of the \emph{rescaled} boundary condition $e^{-\ii \tilde{k}_\beta \tilde{x}^\beta}\!\nes\times\eqref{evan2}$ restricted to $N$ ``deepest-penetrating" surface wave modes, namely 
\begin{equation}\label{evan3}
\sum_{n=1}^{N} a_n\, \btaun \big|_{\mcSd} =\: 0, \qquad 
a_n = e^{\ii\tilde{\kappa}_n\mathfrak{d}}\,\eta_n,
\end{equation}
where 
\begin{equation}
\btaun(\bx) \,=\, G(\bx)\, \bnu(\bx)\cdot\nabla_{\!\bkn}\nes \tilde{\phi}_n(\bx) \: e^{\ii\es\tilde{\kappa}_n (\tilde{x}^\alpha-\mathfrak{d})} 
\end{equation}
denotes the normalized surface traction due to $n$th Bloch mode, and the normalization introduced via~$a_n$ prevents ill-conditioning of the ensuing eigenproblem due to possible disparity in the magnitudes of~$e^{\ii\tilde{\kappa}_n\mathfrak{d}}$ ($n\shh 1,2,\ldots$) for~$\tilde{\kappa}_n$ with large imaginary parts. In this setting, boundary condition~\eqref{evan3} over an undulated surface cut can be parsed with the aid of~\eqref{s7} as 
\begin{equation}\label{evan4}
\sum_{n=1}^{N} a_n\, \btaun(\bx)=\: 0 
\qquad \text{for} \quad \bx \hh=\hh \zeta\tba_\beta +\big(\mathfrak{d} + f_p(\zeta)\big)\hh\tba_\alpha,  \quad \zeta \in \tilde{\Upsilon}_{\!\mcS}\psup, \quad p=\overline{1,P}.  
\end{equation}

\begin{remark} \label{rem3}
We note that both $\bx-\zeta\tba_\beta$ and $\btaun(\bx) = \btaun(\zeta\tba_\beta+(\bx-\zeta\tba_\beta))$ in~\eqref{evan4} are $\tilde{\Upsilon}_{\!\mcS}$-periodic functions of~$\zeta$, which motivates introduction of the rescaled boundary condition $\eqref{evan3} \hh:=\hh e^{-\ii \tilde{k}_\beta \tilde{x}^\beta}\!\nes\times\eqref{evan2}$.    
\end{remark}

With~$a_n$ ($n\shh\overline{1,N}$) solving~\eqref{evan3} at hand, the reduced-order model (ROM) of a surface Bloch wave obtained by restricting~\eqref{evan1} to~$N$ deepest-penetrating modes can be written as 
\begin{equation}\label{evan1rom}
{\sf u}(\bx) \::=\:\Big(\sum_{n=1}^N a_n\hh \tilde{\phi}_n(\bx) \, e^{\ii\es \tilde{\kappa}_n (\tilde{x}^\alpha-\mathfrak{d})}\Big) \hh e^{\ii \nes \tilde{k}_\beta \tilde{x}^\beta}. 
\end{equation}

On introducing an auxiliary set of support functions  
\begin{equation} \label{heavis}
H\psup(\zeta) = \left\{
\begin{array}{ll}
1, & \zeta\in \tilde{\Upsilon}_{\!\mcS}\psup\\
0, & \zeta\notin \tilde{\Upsilon}_{\!\mcS}\psup
\end{array} \right., \qquad p=\overline{1,P}
\end{equation}
boundary conditions \eqref{evan4} can be rewritten over the unit cell of in-plane SB wave periodicity as 
\begin{equation}\label{evan5}
H\psup(\zeta)\sum_{n=1}^{N} a_n\, \btaun(\bx) =\: 0 
\quad \text{for} \quad \bx = \zeta\tba_\beta +\big(\mathfrak{d} + f_p(\zeta)\big)\hh\tba_\alpha, \quad \zeta \in \tilde{\Upsilon}_{\!\mcS}, \quad p=\overline{1,P}. 
\end{equation}

We are now in position to expand~\eqref{evan5} in Fourier series with respect to~$\zeta$ for each~$p$. We proceed by rewriting~\eqref{evan5} as 
\begin{equation}\label{evan7}
H\psup(\zeta)\sum_{n=1}^{N} a_n\, \hat{\tau}_n\psup(\zeta) \:=\: 0 
\quad \text{for} \quad \zeta \in \tilde{\Upsilon}_{\mcS}, \quad p=\overline{1,P}   
\end{equation}
where 
\[
\hat{g}\psup(\zeta) \hh=\hh g(\boldsymbol{x}(\zeta)) \quad \text{for} \quad \bx(\zeta) \hh=\hh \zeta\hh\tba_\beta +\big(\mathfrak{d} + f_p(\zeta)\big)\hh\tba_\alpha.
\]
Since $\tilde{\Upsilon}_\mathcal{S}=\{\zeta: -\tfrac{1}{2}<\nes\zeta\nes<\tfrac{1}{2}\}$, the left-hand side of~\eqref{evan7} can be expanded in Fourier series as 
\begin{equation} \label{fields2}
\begin{aligned}
H\psup(\zeta)\sum_{n=1}^{N} a_n\, \hat{\tau}_n\psup(\zeta) &~= 
\sum_{m=-\infty}^{\infty} \bigg\{ \sum_{n=1}^{N} a_n \,\hat{\psi}_{mn}\psup \bigg\} e^{{i} 2 \pi m \zeta }, \\ 
\hat{\psi}_{mn}\psup &~= \int_{-\frac{1}{2}}^{\frac{1}{2}} H\psup(\zeta)\,\hat{\tau}_n\psup(\zeta) \, e^{-{i} 2 \pi m \hh \zeta} \dd\zeta.
\end{aligned}
\end{equation}
This allows us to enforce the boundary condition~\eqref{evan7} over the leading $2M\!+\nes 1$ Fourier terms as 
\begin{equation} \label{syst1}
\sum_{n=1}^{N} \hat{\psi}_{mn}\psup \, a_n  = 0  \quad \text{for}\quad 
m=\overline{-M,M} ~~~\text{and}~~~  p=\overline{1,P}.
\end{equation}
Recalling~\eqref{evan1}, algebraic system~\eqref{syst1} can be rewritten as a (nonlinear) matrix eigenvalue problem 
\begin{equation}\label{matrix1}
\boldsymbol{\Uppsi}\nes(k_e) \hh \boldsymbol{a} \:=\: \bzero 
\end{equation}
where $\boldsymbol{\Uppsi}$ is a complex-valued $P\hh (2M\!+\nes 1) \times N$ matrix collecting $\hat{\psi}_{mn}\psup$; eigenvalue $\,k_e$ is a prescribed wavenumber in direction~$\be$ featured by QEP~\eqref{QEP-surf3}, and $\ba\shh\ba(k_e)\in\mathbb{C}^N$ is the vector of SB wave amplitudes $a_n$. In the sequel, we refer to~\eqref{matrix1} as the \emph{surface-wave eigenproblem} and assume
\begin{equation} \label{over2D}
P\hh (2M\!+\nes 1) \geqslant N, 
\end{equation}
which caters for (but does not guarantee) having an overdetermined system. 

To expose the wavenumber~$k_e$ (the surface-wave eigenvalue) and the companion vector of QEP modal amplitudes $\ba$ (the surface-wave eigenvector) that describe the SB wave propagating along~$\mcSd$, we next consider the singular value decomposition (SVD) of~$\boldsymbol{\Uppsi}$, namely  
\begin{equation} \label{SVD}
\boldsymbol{\Uppsi}(k_e) \:=\: \boldsymbol{U} \boldsymbol{\Lambda}\hh \overline{\boldsymbol{V}}
\end{equation}
where~$\boldsymbol{U}$ and~$\boldsymbol{V}$ are the unitary matrices of the left and right singular vectors, and $\boldsymbol{\Lambda}=\text{diag}\{\sigma_1,\sigma_2,\ldots\sigma_{N}\}$ compiles the (real, non-negative) singular values arranged in a descending order. In this setting, \eqref{matrix1} permits a non-trivial solution for $k_e=k_e^\star$ where the condition number of~$\boldsymbol{\Uppsi}$ becomes unbounded, i.e.  
\begin{equation} \label{cond1}
C_{\boldsymbol{\Uppsi}}(k_e) := \frac{\sigma_1}{\sigma_{N}} \to \infty \quad \text{as} \quad k_e \to k_e^\star.  
\end{equation}
By virtue of~\eqref{cond1} the surface Bloch wave(s), propagating at frequency~$\omega$ in direction~$\be$ along~$\mcSd$, can be computed from the quadratic eigenspectrum of~\eqref{QEP-surf3} -- written over~$\tilde{Y}$ -- via a line search seeking the peak location(s) $k_e^\star$ of~$C_{\boldsymbol{\Uppsi}}  (k_e)$. On recalling~\eqref{evan1rom}, for each~$k_e^\star$ one obtains a ROM of the SB wave as   
\begin{equation}\label{evan9}
{\sf u}^{\nex\star}\nex(\bx) \:=\, \Big(\sum_{n=1}^{N} a_n^{\nes\star}\hh \tilde{\phi}_n(\bx) \, e^{\ii\es \tilde{\kappa}_n(\tilde{x}^\alpha - \mathfrak{d})}\Big)\hh e^{\ii \tilde{k}_\beta \tilde{x}^\beta}, 
\end{equation}
where $a_n^\star$ are the components of $\ba(k_e^\star)$, and $\{\tilde{\kappa}_n,\tilde{\phi}_n\}_{n=1}^{N}$ solve QEP~\eqref{QEP-surf3} with $\tilde{k}_\beta\shh k_e^\star$. On denoting by~$\bv_{\!N}^\star$ the $N$th right eigenvector of $\boldsymbol{\Uppsi}(k_e^\star)$ i.e.~the $N$th column of~$\bV\nes(k_e^\star)$, from~\eqref{evan3}, \eqref{matrix1} and~\eqref{SVD} we obtain the eigenvector of modal amplitudes as  
\begin{equation} \label{eta-chi}
\ba(k_e^\star) \:=\:  \bv_{\!N}^\star 
\end{equation}
and adopt the normalization $\|\ba\|=1$.

\paragraph{Reduced orthogonal basis for non-bijective undulations.} \label{appen-nonbi}

\noindent For non-bijective undulations ($P\nes>\nes1$) featuring ``small'' supports $\tilde{\Upsilon}_{\!\mcS}\psup$ in that $|\tilde{\Upsilon}_{\!\mcS}\psup|\nes\ll|\tilde{\Upsilon}_{\!\mcS}|$, the Fourier series expansion in Sec.~\ref{SBW}\ref{SWEP} may itself result in having an ill-conditioned matrix~$\boldsymbol{\Uppsi}(k_e)$ for $k_e$ away from $k_e^\star$, and so jeopardize the analysis. Qualitatively speaking, the problem arises from the fact that the function behavior over small segments could be reasonably well approximated by \emph{multiple} combinations of Fourier, and so Bloch, terms. To tackle the problem, it is useful to reformulate the matrix eigenvalue problem~\eqref{matrix1} by resorting to a set of reduced function bases that are orthogonal not only on $\tilde{\Upsilon}_{\!\mcS}$, but also on $\tilde{\Upsilon}_{\!\mcS}\psup$. This is accomplished by the introduction of discrete prolate spheroidal wave (DPSW) functions $\chi_m(\theta;\alpha)$ \cite{Slepian78} that carry the property
\begin{equation}\label{double-per0} 
\int_{-\alpha}^\alpha \chi_m(\theta;\alpha) \, \overline{\chi_n} (\theta;\alpha) \,\text{d}\theta \:=\: \delta_{mn} \,\lambda_m,  \qquad m,n=\overline{-M,M} 
\end{equation}
where $0\nes<\nes\lambda_m \nes<\nes 1$ are the eigenvalues of the prolate matrix~\cite{Slepian78,Varah1993}, constructed from $\{e^{i m \theta} \}_{m=-M}^{M}$ restricted to $(-\alpha,\alpha)$. Clearly, a DPSW function whose eigenvalue~$\lambda_m$ is close to~1 (resp.~0) will have most of its energy localized within $(-\alpha,\alpha)$ (resp. $(-\pi,\pi)\backslash\hh[-\alpha,\alpha]$). For brevity of exposition, we examine the treatment of non-bijective undulations via DPSW expansion in Appendix~B, ESM. 

\subsection{Corroborating pilot estimates} \label{corrob}

The foregoing framework builds upon an implicit tenet that the eigenspectrum of QEP~\eqref{QEP-surf3} is such that the traces $\big\{G\hh \bnu\cdot\nabla_{\!\bkn} \nes \tilde{\phi}_n \big|_{\mcSd}\big\}_{n=1}^{\infty}$ provide a \emph{complete basis} for describing the $\tilde{\Upsilon}_{\!\mcS}$-periodic tractions on~$\mcSd$. While an in-depth analysis of this topic is beyond the scope of our study, we submit a simple example in support of such an idea. 

Consider a ``toy problem'' featuring an orthogonal Bravais lattice $\bR\subset\mathbb{R}^2$ and a straight surface cut in the $\be_1$-direction, in which case~\eqref{QEP-surf3} degenerates to~\eqref{QEP-surf2} with $\tilde{Y}\mapsto Y$. For this configuration, the  (modal) surface traction components in~\eqref{evan3} read
\begin{equation} \label{corrob1}
\tau_n \big|_{\mcS} \,=\, G\hh \bnu\cdot\nabla_{\!\bkn} \nes {\phi}_n \big|_{\mcS} \,=\,  -G\hh \be_2\cdot
\Big[ \nabla{\phi}_n +\ii\hh \big({\kappa}_n\hh \be_2 +  k_1\hh\be_1\big) {\phi}_n\Big]\hh \Big|_{\mcS} \,=\,
-G\hh \Big[ \frac{\partial\phi_n}{\partial\xi_2} +\ii\hh {\kappa}_n\hh  {\phi}_n\Big]\hh \Big|_{\mcS}. 
\end{equation} 
Given the fact that $\sum_{m=1}^{\infty}\tau_n \big|_{\mcS}$ is to be expanded in Fourier series with respect to~$\xi_1$ over $(-\tfrac{1}{2},\tfrac{1}{2})$, our objective is expose the directional variations of~$\phi_n$ for large~$n$. Under the assumptions of Theorem~\ref{thm1}, infinity is the only possible accumulation point of~$\tilde{\kappa}_n$. Thus, our task reduces to investigating~\eqref{QEP-surf2} for large~$|\kappa_n|$, which motivates introduction of the scaling parameter 
\begin{equation} \label{corrob2}
\eps \;=\; \eps(n) \;:=\; \frac{|k_1|}{|\kappa_n|} \;=\; o(1).   
\end{equation}
As demonstrated in Appendix~C(a) via~\eqref{QEP-surf2} with $\psi=\phi_n$ and $\|\phi_n\|_{L^2(Y)}=1$, one finds that 
\begin{equation} \label{corrob3}
\|\nabla \phi_n\|_{L^2_p(Y)} 
~>~ c \, |\kappa_n|, \qquad 0 ~<~ c ~\simeq~ \bigg(8+\frac{G_{\inf}}{G_{\sup}}\bigg)^{\frac{1}{2}} \!-\, 2\hh \sqrt{2} 
\end{equation}
for sufficiently small~$\eps$. Due to the particular structure of~\eqref{QEP-surf2}, however, the last result also implies via a simple perturbation argument that the gradient growth with~$|\tilde{\kappa}_n|$ in~\eqref{corrob3} is \emph{markedly anisotropic} in that 
\begin{equation} \label{corrob4}
\|\be_2\cdot\!\nabla\phi_n\|_{L^2_p(Y)}  ~=~ O \big(\eps \, \|\be_1\cdot\!\nabla\phi_n\|_{L^2_p(Y)} \big)
~=~ O \big(\eps\, \|\nabla\phi_n\|_{L^2_p(Y)}  \big),    
\end{equation}
see Appendix~C(a) for details. Thus for sufficiently large~$|\kappa_n|$, one expects the distributions of~$\phi_n$ to exhibit a \emph{corrugated pattern} with $\|\partial\phi_n/\partial\xi_1\|_{L_p^2(Y)} \gg \|\partial\phi_n/\partial\xi_2\|_{L_p^2(Y)}$, whose oscillations in the (cut) $\xi_1$-direction resemble that of the \emph{Fourier series}. Here, the corrugated behavior of~${\phi}_n$ helps the traces $\big\{G\hh \bnu\cdot\nabla_{\!\bkn} \nes {\phi}_n \big|_{\mcSd}\big\}_{n=1}^{\infty}$ maintain their Fourier-like character \emph{irrespective} of the depth of cut. 

On redefining the small parameter in~\eqref{corrob2} as 
\[
\eps \,=\, \frac{|k_e|}{|\tilde{\kappa}_n|},
\]
we are now in position to generalize the above results by considering a generic Bravais lattice. 

\begin{propo}
For sufficiently large~$|\tilde{\kappa}_n|$,  quadratic eigenfunctions of~\eqref{QEP-surf3} exhibit the behavior 
\begin{equation} \label{corrob5}
\|\nabla\tilde{\phi}_n\|_{L_p^2(\tilde{Y})} ~>~ \lambda\minsub^{1/2} \, c \, |\tilde{\kappa}_n|,  \qquad 0 ~<~ c ~\simeq~   \bigg(8 + \frac{G_{\inf} \tilde{\Gamma}^{\alpha\alpha}}{G_{\sup} \, \tilde{\Gamma}\maxsup}\bigg)^{\frac{1}{2}} - 2\sqrt{2}, 
\end{equation}
where $\tilde{\lambda}\minsub>0$ is the smallest eigenvalue of the contravariant metric tensor $\tilde{\boldsymbol{\Gamma}}$ with components $\tilde{\Gamma}^{ij}\!=\tbb^i\!\cdot\nes\tbb^j$, and $\tilde{\Gamma}\maxsup = \max_{i,j\in\overline{1,2}}|\tilde{\Gamma}^{ij}|$.  Further the cumulative growth in~\eqref{corrob5} is anisotropic in that 
\begin{equation} \label{corrob7}
\|\tbb^\alpha \!\cdot \nabla\tilde{\phi}_n\|_{L_p^2(\tilde{Y})} ~=~ O\big(\eps\, \|\tba_\beta\!\cdot \nabla\tilde{\phi}_n\|_{L_p^2(\tilde{Y})}\big) \quad~   
\text{where}~~\tbb^\alpha\!\!\perp\mathcal{S}  ~~~\text{and}~~ \tba_\beta \parallel\mathcal{S}. 
\end{equation}
See Appendix~C(b) for the proof of~\eqref{corrob5} and justification of~\eqref{corrob7}.
\end{propo}

In physical terms,  \eqref{corrob7} states that the eigenfunction gradient in the reciprocal lattice direction~$\tbb^\alpha$ -- orthogonal to the cut plane -- is small relative to that in the (in-plane) propagation direction $\be = \tba_\beta/\|\tba_\beta\|$. This feature is exemplified in Appendix~D(a), ESM by the plots of~$\tilde{\phi}_n(\bx)$ ($n=\overline{1,25}$) supported on  both orthogonal and skew Bravais lattices. In concluding this section, we note that~\eqref{corrob5} and~\eqref{corrob7}  degenerate respectively to~\eqref{corrob3} and~\eqref{corrob4} by letting $\tilde{\Gamma}^{ij} = \delta_{ij}$, $\,\tba_\beta \nes= \be_1$, and $\hh\tbb^\alpha \nes= \be_2$.  In Section~\ref{results}, we shall further illustrate the above norm estimates via numerical examples. 

\section{Synthesis}

\noindent To provide a succinct perspective of the foregoing developments, for a given \emph{rational}, undulated cut surface~$\mcS_\delta$ (see \eqref{bzeta}--\eqref{s1}, and~\eqref{s7}) we shall denote by $\Omega_{\mcS_\delta}\!\subset\mathbb{R}^2$ the \emph{semi-infinite domain} occupying the $-\bnu_{\nesx\mcS}$ side of~$\mcS_\delta$. For brevity, we will use this notation for both undulated and planar cuts, with an implicit understanding that $\Omega_{\mcS_\delta}\nes\mapsto\Omega_{\mcS}$ in the latter case. With such definition, our aim is to understand the behavior of Bloch wave-like boundary layers governed by 
\begin{equation}\label{PDE2}
\begin{aligned}
 \nabla \sip (G(\bx) \nabla u) + \omega^2\rho(\bx)\hh u \:=\: 0, & \quad \bx \in \Omega_{\mcS_\delta} \\ 
\bnu_{\!\mcSd} \cdot (G(\bx) \nabla u)|_{\mcSd} = 0\; & 
\end{aligned}~,
\end{equation}
where~$G(\bx)$ and~$\rho(\bx)$ are the coefficients of an unbounded ``mother'' periodic medium specified by~\eqref{brav}--\eqref{PDE1}, and $\bnu_{\!\mcSd}$ is the unit outward normal on~$\mcS_\delta$. To guarantee a well-posed boundary value problem, \eqref{PDE2} must in principle be endowed with the radiation condition at infinity, a subject that is beyond the scope of this study. This deficiency is, however, immaterial to our investigation of the surface-wave eigenstates that could be interpreted as a scalar-wave, periodic-medium counterpart of the Rayleigh waves in homogeneous (or layered) elastic media. 

For a rational surface cut~$\mcS$ (see Definition~\ref{def1}), ROM~\eqref{evan9} of the boundary layer is computed by (i) identifying the unit ``multi-cell'' of medium periodicity~\eqref{multiY} that accounts for a geometric interplay between the mother Bravais lattice~$\bR$ and orientation of the surface cut~$\mathcal{S}$; (ii) solving the affiliated QEP~\eqref{QEP-surf3} whose truncated eigenspectrum is used as a basis for the ROM; and (iii) solving a finite-dimensional, surface-wave eigenproblem~\eqref{matrix1} synthesizing the homogeneous boundary condition in~\eqref{PDE2}. For a given (rational) direction~$\be$ of an SB wave and frequency~$\omega$, a nontrivial solution of~\eqref{matrix1} yields the surface-wave wavenumber(s) $k_e^\star$ and companion amplitudes~$a_n^\star$ of the QEP eigenmodes contributing to ROM~\eqref{evan9}. 

In physical terms, eigenvalues $\tilde{k}_\alpha \nes := \tilde{\kappa}_n$ of QEP~\eqref{QEP-surf3} -- featured by the ``modulation'' terms $e^{\ii\es \tilde{\kappa}_n (\tilde{x}^\alpha-\mathfrak{d})}$ in~\eqref{evan9} and so specifying the variation of wave motion in  direction orthogonal to~$\mathcal{S}$ -- are shown to be complex-valued, giving rise to the generation of evanescent boundary layers. As demonstrated in the next section, this feature allows us to explicitly compute the power flow density, i.e. energetic relevance, of a surface Bloch wave due to~\eqref{evan9}. 

\begin{figure}[ht] 
\centering
\includegraphics[width=0.65\linewidth]{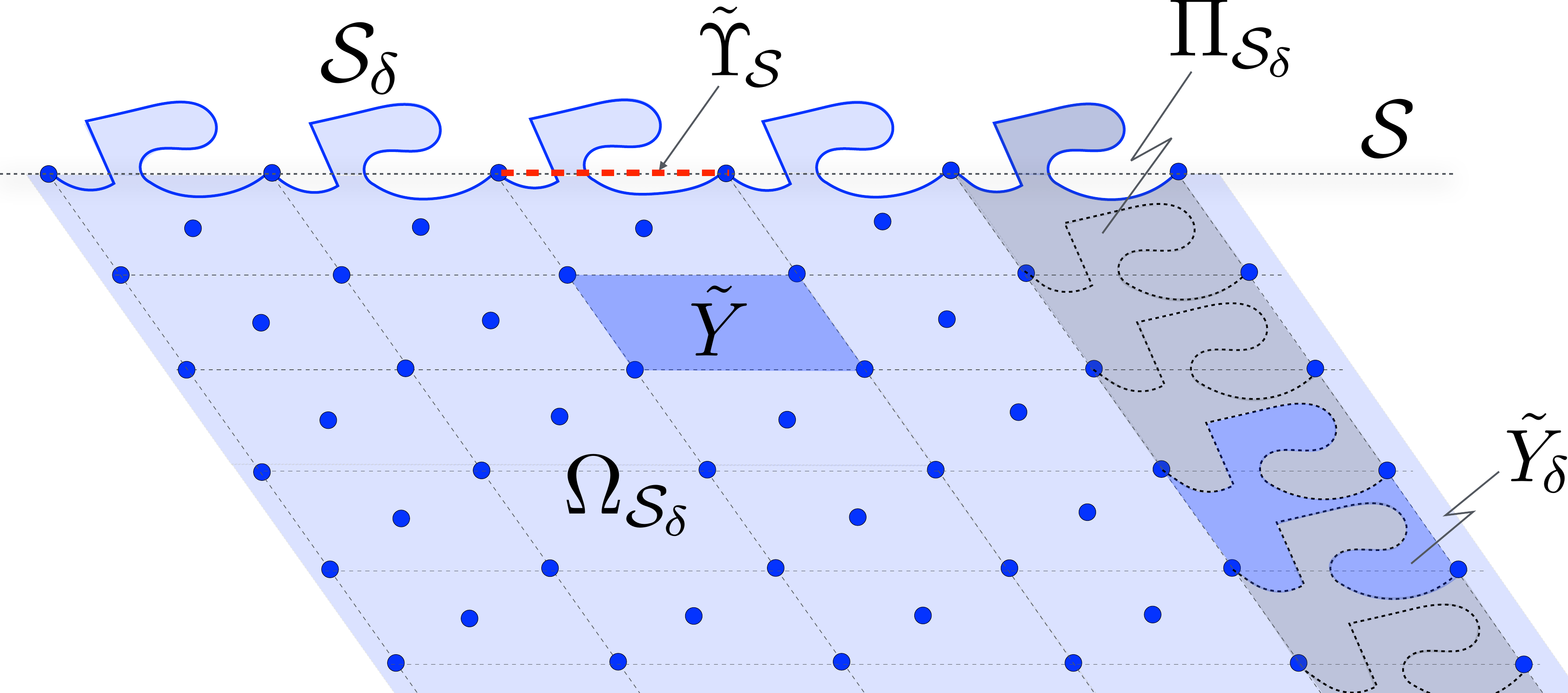}
\caption{Unit prism of periodicity, $\Pi_{\nes\mcS_{\nes\delta}}$, of the semi-infinite domain $\Omega_{\mcS_\delta}$ and re-tailored unit cell of periodicity $\tilde{Y}_{\nesx\delta}$ (with $|\tilde{Y}_{\nesx\delta}|=|\tilde{Y}|$) catering for the surface undulation~$\mcS_{\delta}$.} \label{prism} 
\end{figure}

Going forward, it is also useful to interpret the semi-infinite domain~$\Omega_{\mcS_\delta}\!\subset\mathbb{R}^2$ occupied by the mother periodic medium as a structure that is itself periodic in~$\mathbb{R}$, more specifically in the plane~$\mcS$ of a rational surface cut, see~\eqref{bzeta}--\eqref{s1}. In this vein, one can identify the \emph{unit prism of periodicity} of the ``half-space''~$\Omega_{\mcS_\delta}$ as   
\begin{equation} \label{prism1}
\Pi_{\nes\mcS_{\nes\delta}} = \big\{\bx:\: \bx\in\Omega_{\mcS_\delta}, \; -\tfrac{1}{2}< \bx\nes\cdot\nes\tbb^\beta\!<\tfrac{1}{2}\big\}, 
\end{equation}
whose base~$\tilde{\Upsilon}_{\!\mcS}$ -- namely the cross-section parallel to~$\mcS$ -- is given by~\eqref{s1}, see Fig.~\ref{prism}. 

\subsection{Dispersion of surface Bloch waves} \label{SBdisper}

\noindent Recalling~\eqref{evan9}, we observe that for a given orientation ($\bnu_{\!\mcS}\perp\be$) of the surface cut~$\mcS$, the foregoing analysis results in a \emph{one-dimensional} (1D) Brillouin zone 
\begin{equation} \label{1DBZ}
\tilde{B}_{\nes\mcS} \,=\,  \big\{k_e^\star\nes\in\mathbb{R}:\: -\pi<\nes k_e^\star\nes<\pi \big\},      
\end{equation}
describing the dispersion of surface Bloch waves. Clearly, for 2D problems examined in this study there is a single such Brillouin zone due to the fact that $\be$ is (up to a sign) uniquely determined by the cut orientation. As will be demonstrated shortly, the dispersive characteristics of surface Bloch waves over~$\tilde{B}_{\nes\mcS}$ carry many ``usual'' features such as pass bands and band gaps. In the context of~\eqref{matrix1}, it is worth noting that (for given~$\omega$ and~$\be$) matrix $\boldsymbol{\Uppsi}(k_e)$ may in principle feature repeated eigenvalues~$k_e^\star$, corresponding to a crossing between several (SB wave) dispersion branches. 

\subsection{Power flow and ``skin depth'' of a surface Bloch wave}

\noindent To bring an insight into the power flow of surface Bloch waves, for given $\tilde{k}_\beta=k_e^\star\in\tilde{B}_{\nes\mcS}$ it is instructive to parse~\eqref{evan9} as 
\begin{equation}\label{evan9x}
{\sf u}^{\nex\star}\!(\bx) \;=\; \sum_{n=1}^{N} a_n^\star\hh \hh e^{-\Im(\tilde{\kappa}_n)(\tilde{x}^\alpha-\mathfrak{d})}
\big[\tilde{\phi}_n \, e^{\ii\es \Re(\bkn)\hhs\cdot\hhs(\bx-\mathfrak{d}\hh\tba_\alpha)}\big], \qquad  \bkn= \tilde{\kappa}_n \tbb^\alpha \!+ \tilde{k}_\beta \hh\tbb^\beta, 
\end{equation}
where~$\bx\nes\in\nes\Omega_{\mcS_\delta}$; $\:\tilde{\phi}_n(\bx)$ is $\tilde{Y}$-periodic, and the bracketed term has the character of a (standard) Floquet wave~\eqref{floquet} in an infinite $\tilde{Y}$-periodic medium. 

\begin{defin} \label{def4}
Recalling Fig.~\ref{prism}, we shall denote by~$\tilde{Y}_{\nesx\delta}$ a``re-tailored'' unit cell that (i) shares the support~$\tilde{\Upsilon}_{\!\mcS}$ in the $\mcS$-plane with~$\tilde{Y}$, and (ii) is confined in the $\tba_\alpha$-direction between the undulated surfaces $\mcS_\delta+\ell\hh\tba_\alpha$ and $\mcS_\delta+(\ell\!+\!1)\tba_\alpha$, where $\ell=\mathcal{Z}^+$ and $\mcS_\delta$ is given by~\eqref{s7}. 
\end{defin}

Letting $\dot{u}=\partial u/\partial t$ and~$\bsig = G(\bx) \nabla u$ denote respectively the particle velocity and stress vector, we turn our attention to the Poynting vector~$\boldsymbol{P}(\bx,t)=-\dot{u}(\bx,t)\hh \bsig(\bx,t)$, which specifies the power flow density of wave motion. For time-harmonic motions with implicit factor~$e^{-\ii\omega t}$, its temporal average over the period of oscillations has been shown \cite{willis2016negative} to read $\boldsymbol{P}(\bx) = \tfrac{\omega}{2}\hh\Im[\overline{u(\bx)}\hh\bsig(\bx)]$. In terms of ROM~\eqref{evan9x}, we accordingly have 
\begin{equation}
\boldsymbol{P}^\star\nex(\bx) = \frac{\omega}{2} \hh \Im\big[\hh \overline{{\sf u}^{\nex\star}\nex(\bx)} \hh \bupsig^\star\nex(\bx)\big],  
\end{equation}
where 
\begin{equation}\label{evan9xx}
\bupsig^\star\nex(\bx) \:=\, 
\sum_{n=1}^{N} a_n^\star\hh \hh e^{-\Im(\tilde{\kappa}_n)(\tilde{x}^\alpha-\mathfrak{d})} 
\big[G\hh (\nabla_{\!\bk_n}\!\es \tilde{\phi}_n) \, e^{\ii\es \Re(\bk_n)\hhs\cdot\hhs(\bx-\mathfrak{d}\hh\tba_\alpha)}\big].  
\end{equation}

We are now in position to introduce the power flow density of a surface Bloch wave, averaged over the unit cell of periodicity~$\tilde{Y}_{\nesx\delta}$, as 
\begin{eqnarray} \label{pflow1}
\langle\boldsymbol{P}^\star\rangle \!\!\!\!&:=&\!\!\!\! 
\frac{1}{|\tilde{Y}_{\nesx\delta}|} \int_{\Pi_{\nes\mcS_{\nes\delta}}} \boldsymbol{P}^\star\nesx(\bx) \dd\bx \notag \\
\!\!\!\!&=&\!\!\!\! \frac{\omega}{2\hh|\tilde{Y}_{\nesx\delta}|} \sum_{n=1}^{N}\sum_{m=1}^{N} \int_{\Pi_{\nes\mcS_{\nes\delta}}}     
\Im\Big[\overline{a_n^\star}\hh a_{m}^\star \, e^{-\Im(\tilde{\kappa}_n+\tilde{\kappa}_m)(\tilde{x}^\alpha-\mathfrak{d})} 
\big[\overline{\tilde{\phi}_n}\hh G \,\nabla_{\!\bk_m}\!\tilde{\phi}_m\big]\Big] \dd\bx 
\end{eqnarray}
where the term within inner brackets is $\tilde{Y}$- (and so $\tilde{Y}_{\nesx\delta}$-) periodic, and~$\Pi_{\nes\mcS_{\nes\delta}}$ is the unit prism of periodicity given by~\eqref{prism1}. Recalling Definition~\ref{def4}, we next observe that 
\begin{equation}\label{Pi}
\Pi_{\nes\mcS_{\nes\delta}} = \bigcup_{\ell=0}^\infty \big(\tilde{Y}_{\nesx\delta} + \ell\hh \tba_\alpha\big),  
\end{equation}
where $\tilde{Y}_{\nesx\delta}$ refers to the ``top'' unit cell of~$\Pi_{\nes\mcS_{\nes\delta}}$, bounded in part by~$\mcS_\delta$. Since $\ell\hh\tba_\alpha\!\cdot\tbb^\alpha = \ell$ (no summation over~$\alpha$), from~\eqref{pflow1} and~\eqref{Pi} we obtain 
\begin{eqnarray} \label{pflow2}
\langle\boldsymbol{P}^\star\rangle \!\!\!\!&=&\!\!\!\! 
\frac{\omega}{2\hh|\tilde{Y}_{\nesx\delta}|} \sum_{\ell=0}^\infty\sum_{n=1}^{N}\sum_{m=1}^{N} \int_{\tilde{Y}_\delta}     
\Im\Big[\overline{a_n^\star}\hh a_{m}^\star \, e^{-\Im(\tilde{\kappa}_n+\tilde{\kappa}_m)(\tilde{x}^\alpha\nes-\mathfrak{d}\hh+\hh\ell)} 
\big[\overline{\tilde{\phi}_n}\hh G \,\nabla_{\!\bk_m}\!\tilde{\phi}_m\big]\Big] \dd\bx  \notag \\
 \!\!\!\!&=&\!\!\!\! 
\frac{\omega}{2\hh|\tilde{Y}_{\nesx\delta}|} \sum_{n=1}^{N}\sum_{m=1}^{N} \sum_{\ell=0}^\infty 
e^{-\Im(\tilde{\kappa}_n+\tilde{\kappa}_m)\hh\ell} \int_{\tilde{Y}_\delta}     
\Im\Big[\overline{a_n^\star}\hh a_{m}^\star \, e^{-\Im(\tilde{\kappa}_n+\tilde{\kappa}_m)(\tilde{x}^\alpha-\mathfrak{d})} 
\big[\overline{\tilde{\phi}_n}\hh G \,\nabla_{\!\bk_m}\!\tilde{\phi}_m\big]\Big] \dd\bx \notag \\
 \!\!\!\!&=&\!\!\!\! 
\frac{\omega}{2\hh|\tilde{Y}_{\nesx\delta}|} \sum_{n=1}^{N}\sum_{m=1}^{N}
\frac{1}{1-e^{-\Im(\tilde{\kappa}_n+\tilde{\kappa}_m)}} \int_{\tilde{Y}_\delta}     
\Im\Big[\overline{a_n^\star}\hh a_{m}^\star \, e^{-\Im(\tilde{\kappa}_n+\tilde{\kappa}_m)(\tilde{x}^\alpha-\mathfrak{d})} 
\big[\overline{\tilde{\phi}_n}\hh G \,\nabla_{\!\bk_m}\!\tilde{\phi}_m\big]\Big] \dd\bx. ~\qquad 
\end{eqnarray}
This result allows us to compute the averaged power flow density of a surface Bloch wave, with no truncation of the semi-infinite prism~$\Pi_{\nes\mcS_{\nes\delta}}$, by piecewise integration over the (re-tailored) unit cell~$\tilde{Y}_\delta$. Recalling Fig.~\ref{prism}, we emphasize that $\langle\boldsymbol{P}^\star\rangle$ quantifies the power flow density accrued over the \emph{unit prism} of periodicity $\Pi_{\nes\mcS_{\nes\delta}}$ and scaled by the measure, $|\tilde{Y}_{\nesx\delta}|$, of the \emph{unit cell} of periodicity.

\begin{remark}
The QEP eigenfunctions~$\tilde{\phi}_n$ in~\eqref{pflow2}, interpreted as~$\tilde{Y}$-periodic maps $\mathbb{R}^2\mapsto\mathbb{C}$, are independent of the choice of the free surface undulation $\mcS_\delta$. In fact, the only undulation-dependent items in~\eqref{pflow2} are the modal amplitudes~$a_n^\star$ stemming from the surface-wave eigenproblem~\eqref{matrix1} and the support of integration, $\tilde{Y}_\delta\subset\mathbb{R}^2$. This fact notably caters for an optimal undulation design targeting the manipulation of surface waves, e.g. via controlling their penetration depth (introduced next) or making an appeal to their undulation-dependent dispersion -- featuring for instance surface-wave band gaps. 
\end{remark}

\subsubsection{Skin depth}

\noindent When considering the wave motion in compactly-supported cutouts of periodic media, it may be of interest to estimate the \emph{depth of penetration}, i.e.~skin depth~$\mathfrak{s}$, of surface waves. Result~\eqref{pflow2} motivates us to evaluate this quantity in an energetic sense, as a depth below which the contribution to the averaged power flow density of a surface Bloch wave, $\langle\boldsymbol{P}^\star\rangle$, drops below a given (relative) threshold~$\vartheta$. To this end one may seek a \emph{quantal} skin depth measure $L_\mathfrak{s}$, referring to the number of unit cells, via a simplified condition seeking \emph{the smallest}~$L$ such that 
\begin{equation}\label{skindepth1a}
\sum_{\ell=0}^{L} \sum_{n=1}^N \sum_{m=1}^N e^{-\Im(\tilde{\kappa}_n+\tilde{\kappa}_m)\hh\ell}\: |a_n^\star\hh a_m^\star|
\;>\; (1-\vartheta)\, \sum_{n=1}^N \sum_{m=1}^N \frac{1}{1-e^{-\Im(\tilde{\kappa}_n+\tilde{\kappa}_m)}} \, |a_n^\star\hh a_m^\star|, \qquad \vartheta = o(1),
\end{equation}
which yields the skin depth 
\begin{equation}\label{skindepth1b}
\mathfrak{s} \,=\, \frac{L_{\mathfrak{s}}\shp 1}{\|\tbb^\alpha\|} \,=\, (L_{\mathfrak{s}}\shp 1) \hh \|\tba_\alpha\|. 
\end{equation}

As another avenue toward estimating the skin depth, one may recall~\eqref{s0} and specify~$\mathfrak{s}$ as the normal distance from the cut surface, 
\begin{equation}
\mathfrak{s} \,=\, (\tilde{x}^\alpha\! -\mathfrak{d}) \hh \|\tba_\alpha\| \,:=\, \tilde{s}\hh\es \|\tba_\alpha\|
\end{equation}
where the \emph{nominal amplitude} of a surface Bloch wave becomes negligible relative to its surface value. By virtue of~\eqref{evan9x}, this parameter can be estimated via the condition
\begin{equation} \label{skindepth2b}
\sum_{n=1}^{N} |a_n^\star|\, e^{-\Im(\tilde{\kappa}_n)\hh \tilde{s}} ~=~ \vartheta \sum_{n=1}^{N} |a_n^\star|,  \qquad \vartheta = o(1). 
\end{equation} 
Clearly, both estimates due to~\eqref{skindepth1a} and~\eqref{skindepth2b} take into account the relative amplitudes of individual QEP eigenmodes, the key difference being that they rely on the decay with depth of power flow density and evanescent wave amplitude, respectively. 

\section{Numerical results} \label{results}

\noindent With reference to a Cartesian frame 
endowed with an orthonormal basis~$\be_i$, we consider a periodic medium supported on an orthogonal lattice $\bR\hh\subset\hh\mathbb{R}^2$ featuring the square unit cell
\begin{equation} \label{ex1-cellgeo}
Y \,=\: \big\{\bx=\nes \xi_i\hh \be_i\nes:\: -\tfrac{1}{2}< \xi_i\!<\tfrac{1}{2}\big\},   \qquad 
(G,\rho) \;=\; \left\{ \begin{array}{ll}
\!(5,0.1), ~ & \|\bx\| < 0.2, \\*[1mm]
\!(1,1),   ~ & \text{otherwise} 
\end{array} \right., \quad \bx\in Y.
\end{equation} 

\subsection{Elemental example} \label{elemental}

\begin{figure}[h!] 
\centering{\includegraphics[width=0.98\linewidth]{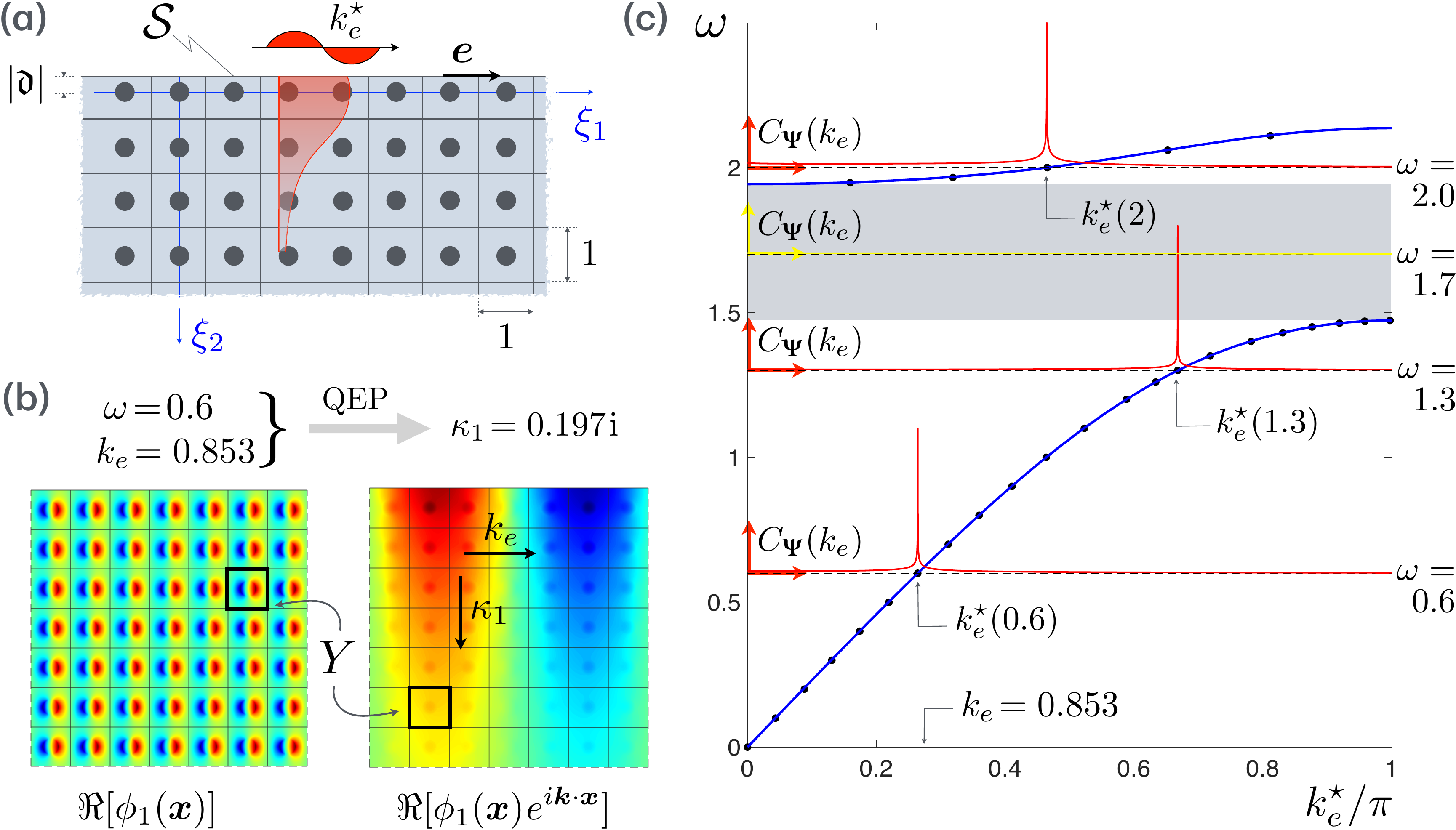}}
\caption{Dispersion of surface Bloch waves: (a) schematics of an evanescent Bloch wave~\eqref{evan1} propagating in direction $\be\nes=\nes\be_1$ along the free surface $\mathcal{S}\nes=\nes\{\bx\!: \xi_2\nes=\nes\mathfrak{d}\}$ of a periodic half-space for some $\mathfrak{d}\!\in\![-0.5,0.5)$; (b) example QEP solution, fundamental mode: $\phi_1(\bx)$ and $\phi_1(\bx)\hh e^{\bk\cdot\bx}$  (real parts) which illustrate the genesis of SB waves; (c) the first two branches of the \emph{surface-wave} dispersion diagram for $\mathfrak{d}\nes=\nes -0.5$: bullets -- spectral data $k_e^\star(\omega)$ computed via QEP~\eqref{QEP-surf2} and condition~\eqref{cond1}, and solid lines -- dispersion of (standard) Floquet-Bloch waves~\eqref{floquet} propagating in direction~$\be_1$ through an \emph{unbounded} periodic medium. In the last panel, example (scaled) distributions of $C_{\boldsymbol{\Uppsi}}  (k_e)$ for $\omega\nes\in\nes\{0.6, 1.3,2\}$ each feature a sharp peak with an $O(10^3)$ magnitude, locating a point on the SB dispersion diagram. By contrast, $C_{\boldsymbol{\Uppsi}}  (k_e)$ for $\omega\!=\!1.7$ (inside the surface-wave band gap, shaded area) is monotonic with an $O(10)$ maximum value.} \label{surdis1} \vspace*{-5mm}
\end{figure}

\noindent As shown in Fig.~\ref{surdis1}(a), the surface cut is made in the $\xi_1$-direction, initially at a depth \mbox{$\mathfrak{d}=-\tfrac{1}{2}$} corresponding to an interface between the two ``rows'' of unit cells. For this configuration, relevant QEP calculations are performed over the unit cell of the periodic medium by setting $\tilde{Y} \mapsto Y$, $\tilde{\phi}_n\mapsto\phi_n$ and $\tilde{\kappa}_n\mapsto\kappa_n$. Accordingly, we seek to reconstruct the SB wave via superposition of QEP eigenstates solving~\eqref{QEP-surf2} that satisfies the traction-free boundary condition on~$\mathcal{S}$. To illustrate the solution of~\eqref{QEP-surf2}, Fig.~\ref{surdis1}(b) plots the first i.e. ``least-evanescent'' eigenfunction~$\phi_1$ (left panel) and the corresponding Bloch eigenmode~$\phi_1 \hh e^{\ii\bk\cdot\bx}$ with $\bk=(k_e,\kappa_1)$ (right panel) in the half-space at frequency $\omega\shh 0.6$ and trial surface wavenumber \mbox{$k_e\shh0.853$}. In unison, these plots highlight the genesis of evanescent Bloch eigenmodes as the building blocks of SB waves.  

\paragraph{Surface wave dispersion.} To compute the dispersion diagram, for each input frequency $\omega$ and a set of trial wavenumbers $k_{e}^{(q)}\!\in[0,\pi]$ $(q\!=\!1,2,\ldots)$, we (i) solve QEP~\eqref{QEP-surf2}; (ii) adopt the ROM~\eqref{evan1rom} with~$N\!=\!10$; (iii) evaluate the condition numbers~$C_{\boldsymbol{\Uppsi}}(k_e^{(q)})$, and (iv) identify the wavenumber $k_e^\star\!=\!k_e^\star(\omega)$ of the surface Bloch wave via criterion~\eqref{cond1}. In Fig.~\ref{surdis1}(c), points on the first two branches of the surface-wave dispersion diagram identified in this way are indicated by bullets. In the search for~$k_e^\star$, by $k_{e}^{(q)}$ $(q\!=\!1,2,\ldots)$ we sample only a small subset of the Brillouin zone, as identified from the value of $k_e^\star(\omega)$ obtained at previously sampled frequency. For transparency, we also plot the (scaled)  distributions $C_{\boldsymbol{\Uppsi}}(k_e)$ over the entire positive half of the Brillouin zone for \mbox{$\omega\!\in\!\{0.6,1.3,1.7,2\}$}. For $\omega\!\in\!\{0.6,1.3,2\}$, these distributions each feature a sharp peak, locating a point on the dispersion diagram via~\eqref{cond1}. By contrast, inside the surface-wave bandgap,  $C_{\boldsymbol{\Uppsi}}(k_e)$ is monotonic with notably smaller maximum value. For completeness, the distributions  of~$C_{\boldsymbol{\Uppsi}}(k_e)$ at 14 sample frequencies inside the first pass band are plotted in Appendix~D(b), ESM.

\paragraph{Verification.} To substantiate the surface-wave dispersion result in Fig.~\ref{surdis1}(c), we note that the periodic medium~\eqref{ex1-cellgeo} is symmetric with respect to the $\xi_2\nes=\nes m\nes-\nes\tfrac{1}{2}$ and $\xi_2\nes=\nes m$ planes, $m\nes\in\nes\mathbb{Z}$. As a result, the Floquet-Bloch wavefield~\eqref{floquet} propagating in the $\xi_1$-direction ($\bk\parallel\!\be_1$) will inherit these symmetries. Recalling~\eqref{homNeumann}, we accordingly find that 
\begin{equation} \label{homNeumann2}
\bnu_{\nesx\mcS} \cdot \big(G(\bx) \nabla u\big)|_{\mcS} \,=\, \left. - G(\bx) \hh \frac{\partial u}{\partial\xi_2}\right|_{\xi_2=\mathfrak{d}} \!\equiv\, 0 \qquad \text{for} ~~ \mathfrak{d}\in\big\{\! -\tfrac{1}{2},0\big\}, 
\end{equation}
namely that the Floquet-Bloch waves in an unbounded periodic medium~\eqref{ex1-cellgeo} generate \emph{zero traction} along the surfaces $\xi_2\nes =\nes \mathfrak{d}$ for $\mathfrak{d}\in\{-\tfrac{1}{2},0\}$. In other words, we encounter a degenerate situation where the SB wave~\eqref{evan1} \emph{coincides} with the Floquet-Bloch wave~\eqref{floquet} in the semi-infinite region $\xi_2\nes>\nes\mathfrak{d}$ and so has the identical dispersion characteristics. This observation is confirmed in  Fig.~\ref{surdis1}(c), where the conventional Floquet-Bloch dispersion diagram is plotted as solid lines. As seen from the display, the two sets of results (surface-wave vs. body-wave dispersion) are indistinguishable for both dispersion branches. 

\subsection{Effect of the depth of cut}

Clearly, in the above configuration the SB waves lose their evanescent character for they do not decay with increasing $\xi_2$ and so have an unbounded skin depth, $\mathfrak{s}\!\to\!\infty$, according to either~\eqref{skindepth1b} or~\eqref{skindepth2b}. With reference to Fig.~\ref{surdis1}, this is reflected in the relevant QEP eigenvalues~$\tilde{\kappa}_n$ having vanishingly small imaginary parts for~$k_e\nes=k_e^\star$. Naturally, such degeneracy disappears for a generic periodic medium or a generic surface cut, e.g. $\mathfrak{d}\notin\{-\tfrac{1}{2},0\}$ or $\be\nes\neq\nes\pm\be_1$ in the above example. To explore the cut-driven parametric variations, we proceed by examining the effect of~$\mathfrak{d}$ first. 

\begin{figure}[h!] 
\centering{\includegraphics[width=0.87\linewidth]{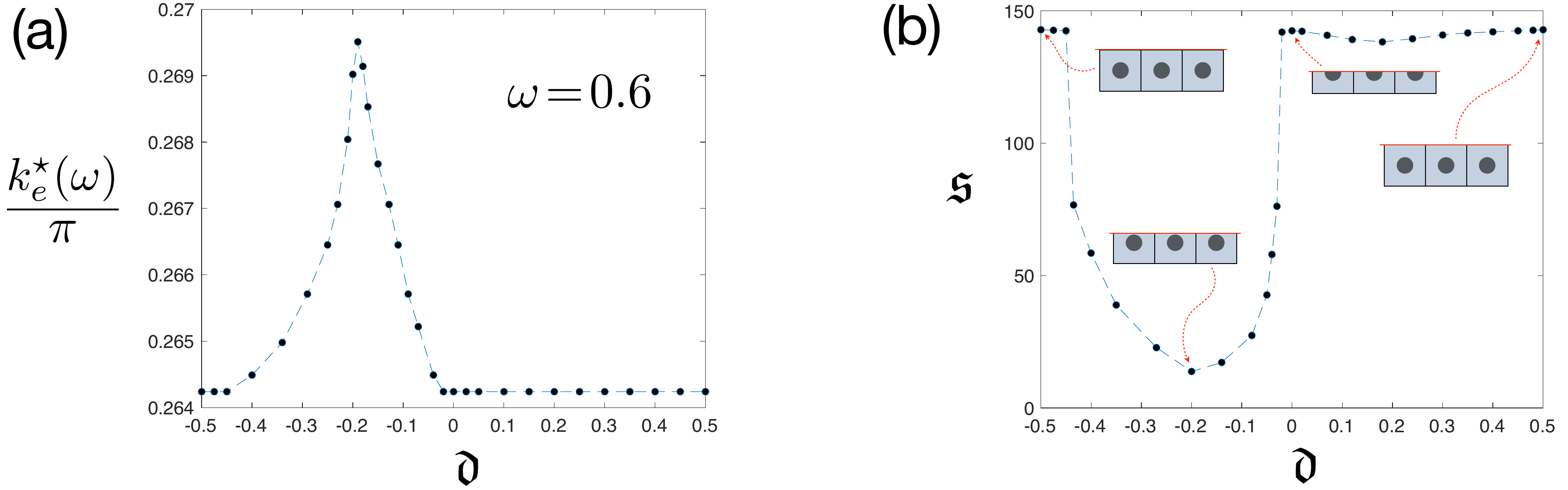}} \vspace*{-1mm}
\caption{Effect of the depth of cut $\mathfrak{d}$ on: (a) surface wavenumber~$k_e^\star$, and (b) skin depth~$\mathfrak{s}$ due to~\eqref{skindepth2b} with $\vartheta = 0.1$.} \label{cut-depth} 
\end{figure}

\paragraph{Skin depth.}  
 As an example, Fig.~\ref{cut-depth} plots the variation of the surface wavenumber~$k_e^\star$ (left panel) and penetration depth~$\mathfrak{s}$ due to~\eqref{skindepth2b} (right panel) versus depth of cut, $\mathfrak{d}$, at frequency $\omega=0.6$. From the figure, one may in particular note that (i) the variations $k_e^\star(\mathfrak{d})$ and $\mathfrak{s}(\mathfrak{d})$ are 1-periodic as governed by the ``height'' of the unit cell measured in direction~$\be_2$, and (ii) the maximum values of $\mathfrak{s}(\mathfrak{d})$ are reached for $\mathfrak{d}\in\{-\tfrac{1}{2},0,\tfrac{1}{2}\}$ as elucidated in Section~\ref{results}\ref{elemental}. 

  \vspace*{-7mm}

\begin{figure}[h!] 
\centering{\includegraphics[width=0.96\linewidth]{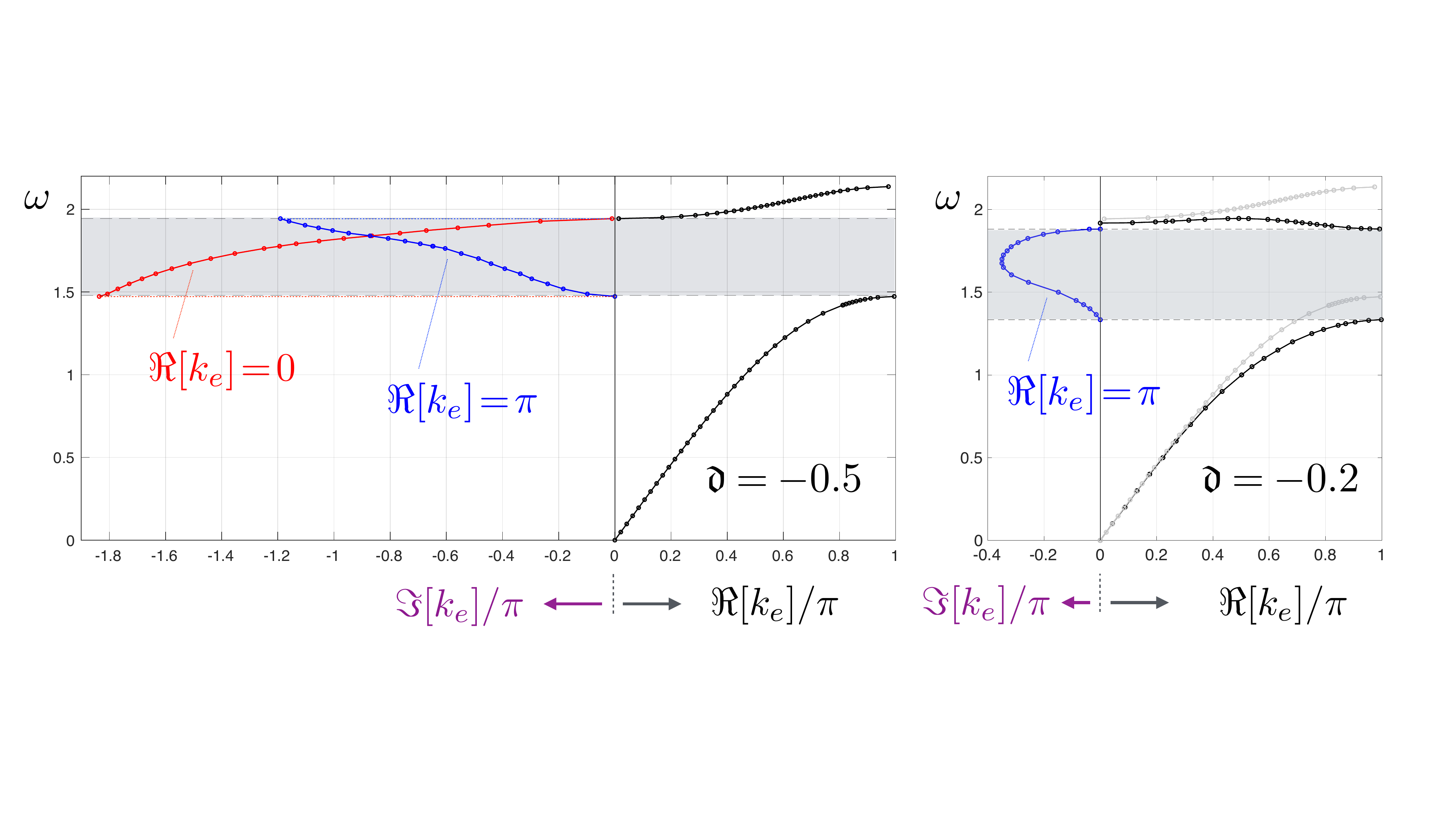}} \vspace*{-1mm}
\caption{Dispersion of SB waves, including behavior inside a band gap: $\mathfrak{d}\shh-0.5$ (left panel), and $\mathfrak{d}\shh-0.2$ (right panel).} \label{dispercut} 
\end{figure}

\paragraph{Dispersion.}  
The effect of the depth of cut on the dispersion of SB waves is illustrated in Fig.~\ref{dispercut}, featuring the results for $\mathfrak{d}\shh-0.5$ and $\mathfrak{d}\shh-0.2$. As seen from the display, changing~$\mathfrak{d}$ from the former to the latter value brings about: (i) lowering and widening of the first band gap, and (ii) flattening of the first two pass bands. To highlight the differences, the right panel (showing the dispersion for $\mathfrak{d}\shh-0.2$) also includes the results for $\mathfrak{d}\shh-0.5$ in light gray. Perhaps more striking is the behavior of SB waves inside the band gap, obtained by letting either $k_e\shh \ii\hh\alpha$ or $k_e\shh \pi+\ii\hh\alpha$ in~\eqref{matrix1} and then seeking the value of~$\alpha\in\mathbb{R}$ that yields the peak condition number according to~\eqref{cond1}. For $\mathfrak{d}\shh-0.5$, one observes that the first pass band ``ends'' at $k_e\shh\pi$, while the second branch ``begins'' at $k_e\shh0$. It is thus no surprise that this configuration yields \emph{two evanescent branches} of the SB waves inside the band gap, affiliated respectively with the roots featuring $\Re[k_e]\shh 0$ and~$\Re[k_e]\shh \pi$. We confirmed this result using the verification approach examined in Section~\ref{results}\ref{elemental} (results not shown for brevity). For $\mathfrak{d}\shh-0.2$, on the other hand, both exceptional points (i.e. edges of the band gap) occur at $k_e\shh\pi$, resulting in a single ``belly-shaped'' evanescent branch affiliated with~$\Re[k_e]\shh \pi$. 

\begin{remark}
For clarity we note that the pass bands in the left panel of Fig.~\ref{dispercut}, computed for $\mathfrak{d}\shh-0.5$, are precisely those featured in Fig.~\ref{surdis1} -- which happen to replicate the dispersive behavior of Floquet-Bloch (body) waves. As seen from the right panel in Fig.~\ref{dispercut}, changing the depth of cut to $\mathfrak{d}\shh-0.2$ causes the dispersion of surface Bloch waves to deviate significantly from their Floquet-Bloch counterpart.    
\end{remark}

\subsection{ROM convergence and behavior of the QEP eigenspectrum} \label{converg} 

\begin{figure}[h!] 
\centering
\includegraphics[width=0.86\linewidth]{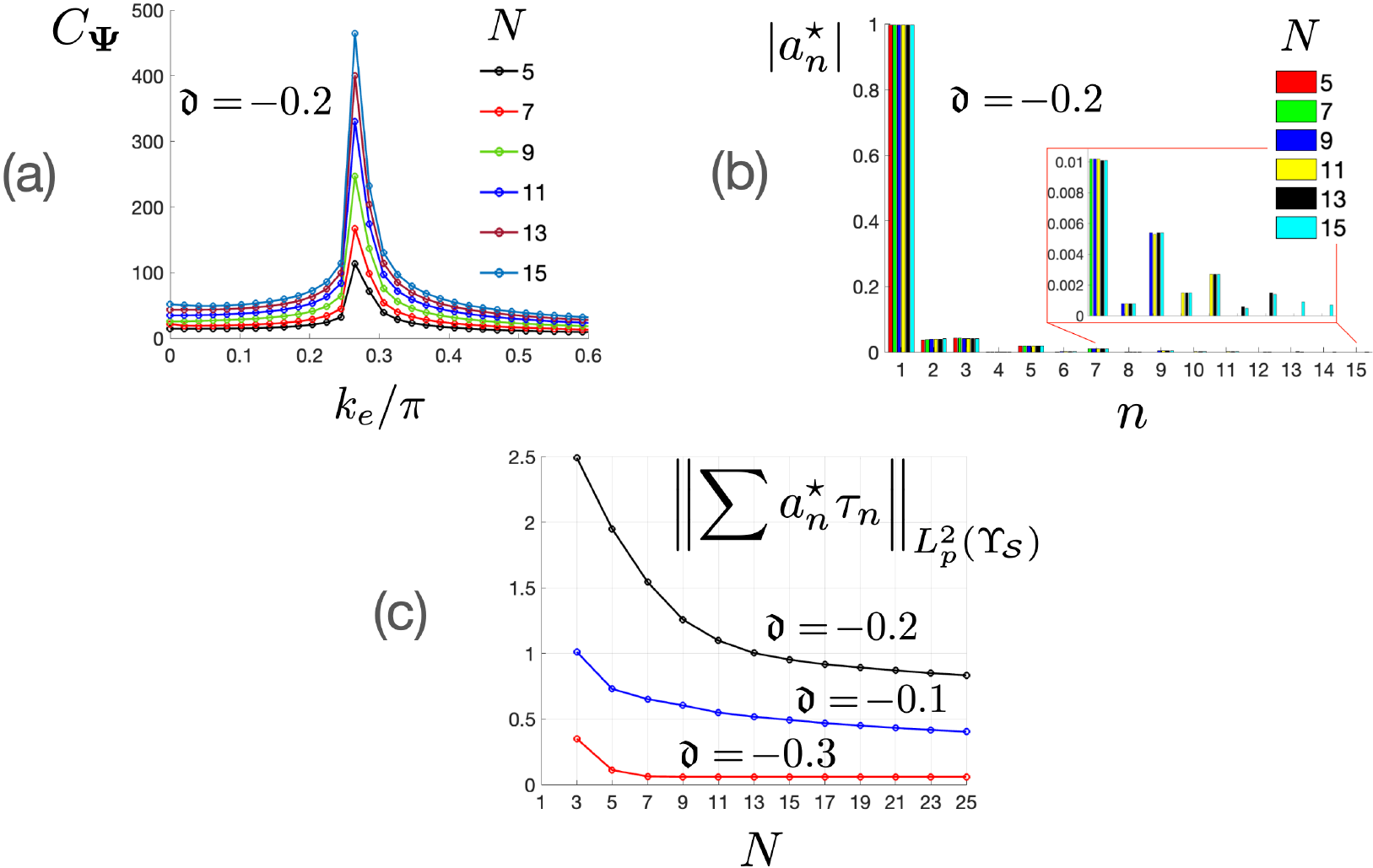} \vspace*{-2mm}
\caption{Numerical convergence of ROM~\eqref{evan9} for $\omega\shh0.6$: (a) condition number, $C_{\boldsymbol{\Uppsi}}(k_e)$, with increasing number~$N$ of the Bloch eigenmodes ($\mathfrak{d}\shh-0.2$); (b) amplitudes~$|a_n^\star|$ ($n=\overline{1,N}$) of the contributing Bloch eigenmodes with increasing~$N$  ($\mathfrak{d}\shh-0.2$), and (c) diminishing norm of the surface traction with~$N\hh$ for $\hh\mathfrak{d}\nes\in\nes\{-0.1,-0.2,-0.3\}$.} \label{convergence} 
\end{figure}

In support of the foregoing results, Fig.~\ref{convergence} illustrates the numerical convergence of ROM~\eqref{evan9} at frequency \mbox{$\omega\shh0.6$}. In particular, Fig.~\ref{convergence}(a) plots the condition number $C_{\boldsymbol{\Uppsi}}(k_e)$ with increasing number~$N$ of the evanescent Bloch eigenmodes for a depth of cut $\mathfrak{d}\shh-0.2$. As expected, $N$ has a major effect on the magnitude of $C_{\boldsymbol{\Uppsi}}$ in that the peak becomes more pronounced with increasing~$N$. Beyond $N=5$, however, the \emph{peak locations} ($k_e\shh k_e^\star$) are visually indistinguishable, suggesting fast convergence of a truncated solution. This is confirmed in Fig.~\ref{convergence}(b) where the "low-mode" amplitudes~$a_n^\star$ ($n\shh\overline{1,5}$) show only a minimal variation for $N\geqslant 5$, while the ``high-mode'' amplitudes~$a_n^\star$ ($n>5$) are negligible irrespective of~$N$. For completeness, Fig.~\ref{convergence}(c) plots the residual $\|\tau\|_{L^2_p(\Upsilon_{\mathcal{S}})}$ in approximating the traction-free boundary condition~\eqref{evan2} versus~$N$ for several depths of cut: $\mathfrak{d}\!\in\!\{-0.1,-0.2,-0.3\}$. Even though the residual diminishes slowly with~$N$ for $\mathfrak{d}\shh-0.2$, this fact does not appear to affect the ROM convergence. 

\begin{figure}[h!] 
\centering
\includegraphics[width=1.0\linewidth]{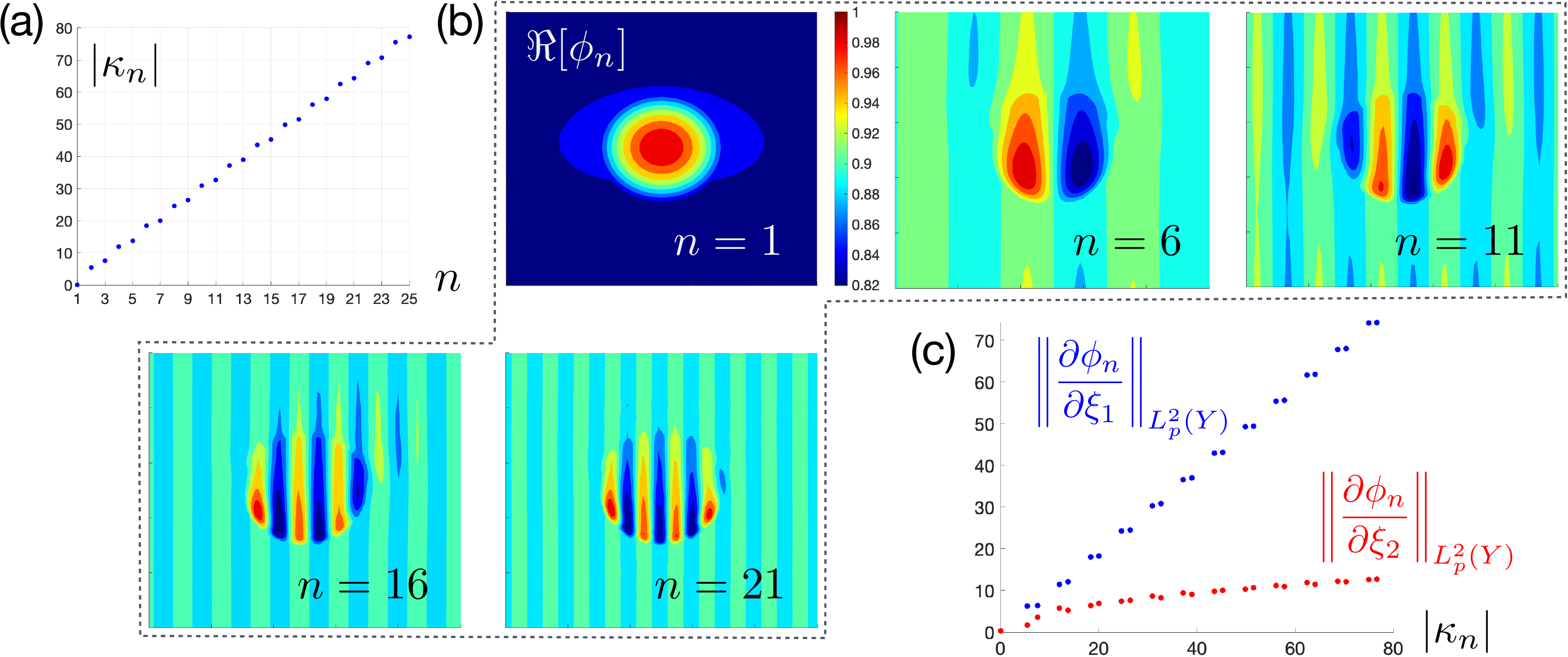} \vspace*{-2mm}
\caption{Behavior of the QEP eigenspectrum at frequency $\omega=0.6$: (a) magnitude of~$\kappa_n$ versus~$n$, (b) selected distributions~$\phi_n(\bx)$ (real parts), and (c) $L_p^2(Y)$-norm of the components of~$\nabla\phi_n$ versus~$|\kappa_n|$.}  \label{completeness} 
\end{figure}

To better understand the above results, Fig.~\ref{completeness}(a) plots~$|\kappa_n|$ for the first 25 eigenmodes of QEP~\eqref{QEP-surf2}. Noting  that all eigenvalues in this example are purely imaginary, we first observe that $\Im[\kappa_n]=|\kappa_n|$ grow approximately linearly with~$n$. In the context of Section~\ref{SBW}\ref{corrob} which exposes the (i) corrugated nature and~(ii) Fourier-like variation in the cut direction of the QEP eigenfunctions for large~$|\kappa_n|$, Fig.~\ref{completeness}(b) plots the variation of sample eigenfunctions~$\phi_n$, while Fig.~\ref{completeness}(c) confirms the anisotropic scaling with~$|\kappa_n|$ of the components of~$\nabla\phi_n$ due to~\eqref{corrob4}. For completeness, spatial variation of the complete set~$\phi_n$ ($n\in\overline{1,25}$) is plotted in Appendix~D(a) (ESM).

\subsection{Effect of surface undulation} \label{num-undul}

\begin{figure}[h!] 
\centering{\includegraphics[width=0.75\linewidth]{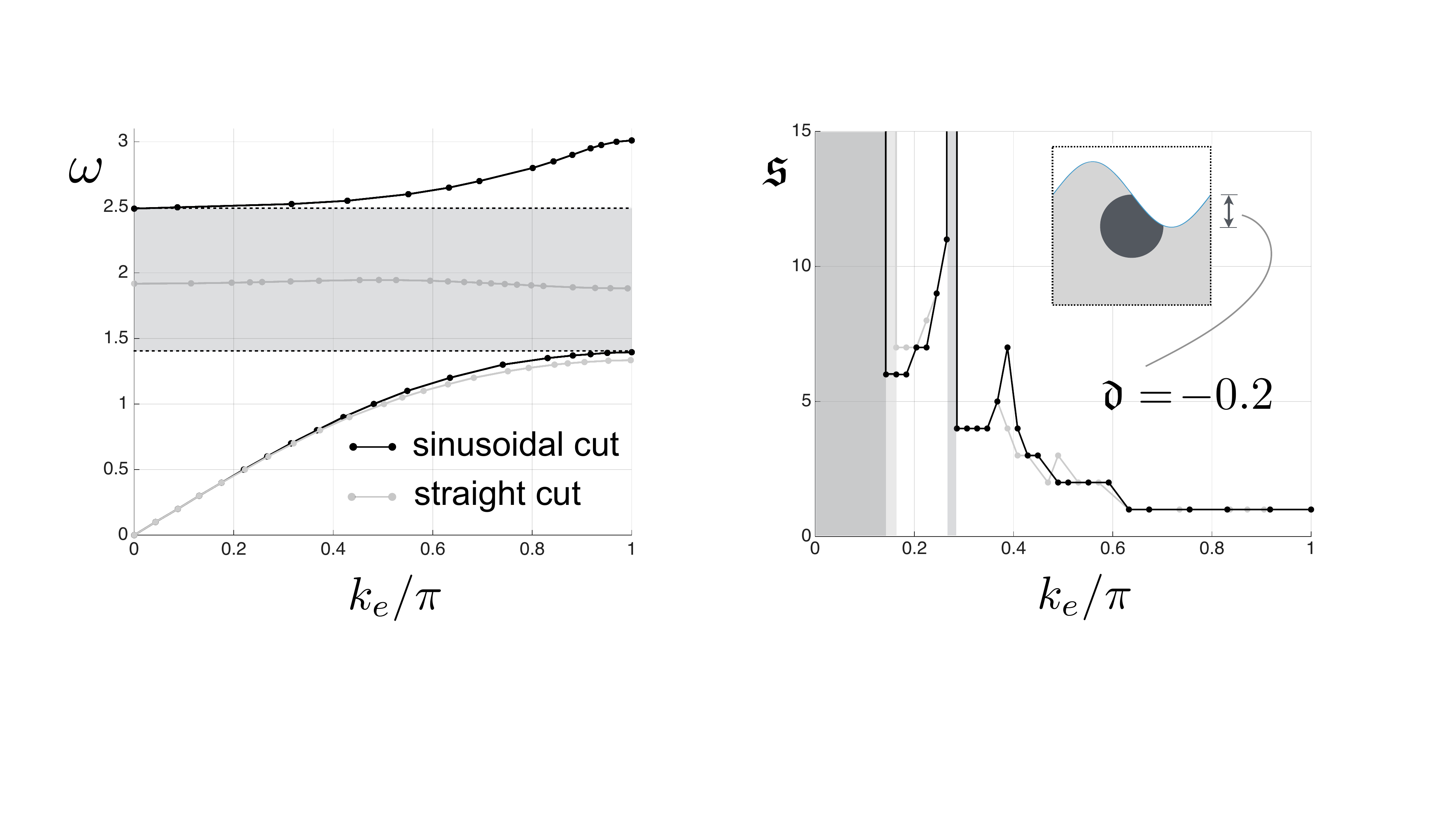}} \vspace*{-1mm}
\caption{Propagation of surface Bloch waves along a ``horizontal'' cut endowed with sinusoidal undulation \mbox{$x_2 = \mathfrak{d}(1+\sin(2\pi x_1))$}: dispersion diagram (left panel), and variation of skin depth vs.~$k_e$due to~\eqref{skindepth1a}--\eqref{skindepth1b} (right panel). As a point of reference, results for the straight cut are plotted alongside in light gray. In the right panel, shaded areas indicate the regions of deep penetration of the boundary layer where $\mathfrak{s}\simeq O(10^2).$} \label{corrug-sine}
\end{figure}

Our next example focuses on the boundary layers developing near an undulated surface of a periodic half-space given by $x_2= \mathfrak{d}(1+\sin(2\pi x_1))$. Letting $\mathfrak{d}\shh-0.2$, the left panel in Fig.~\ref{corrug-sine} plots the first two branches of the SB dispersion diagram, where the previous result for a straight cut ($x_2\shh\mathfrak{d}$) is shown in light gray. From the display, it is seen that the lower edge of the band bap is barely altered by undulation, while its width is nearly doubled. In general, it is foreseen that periodic undulations with (narrow) cantilever-like protrusions may lead to appreciable lowering of the band gap on  account of the local ``cantilever'' resonances. To provide a more complete picture of the boundary layer, the right panel in Fig.~\ref{corrug-sine} plots the variation of the skin depth, $\mathfrak{s}(k_e)$, due to quantal estimate~\eqref{skindepth1a}--\eqref{skindepth1b} for the first branch. The results for both sinusoidal and straight cut are similar and show significant variation of the skin depth with wavenumber (and so frequency). In particular, one may observe the regions (shaded in gray) in the low-wavenumber regime where the skin depth is exceedingly large. This phenomenon is found to occur in situations where the first QEP eigenvalue~$\kappa_1(\omega,k_e)$ solving~\eqref{QEP-surf1} has very small imaginary part. In this way, any boundary layer that features sufficiently small $\Im(\kappa_1)$ and non-trivial modal amplitude~$a_1^\star$ will have significant skin depth. This observation is reflected in the fact that the SB waves for straight and sinusoidal cuts have very similar first dispersion branches and so regions of deep penetration. By contrast, for $k_e>0.6\pi$ the SB waves are in both cases only \emph{one-unit-cell deep}. 

\subsection{Rational surface cut} \label{num-rational}

\begin{figure}[h!] 
\centering{\includegraphics[width=0.88\linewidth]{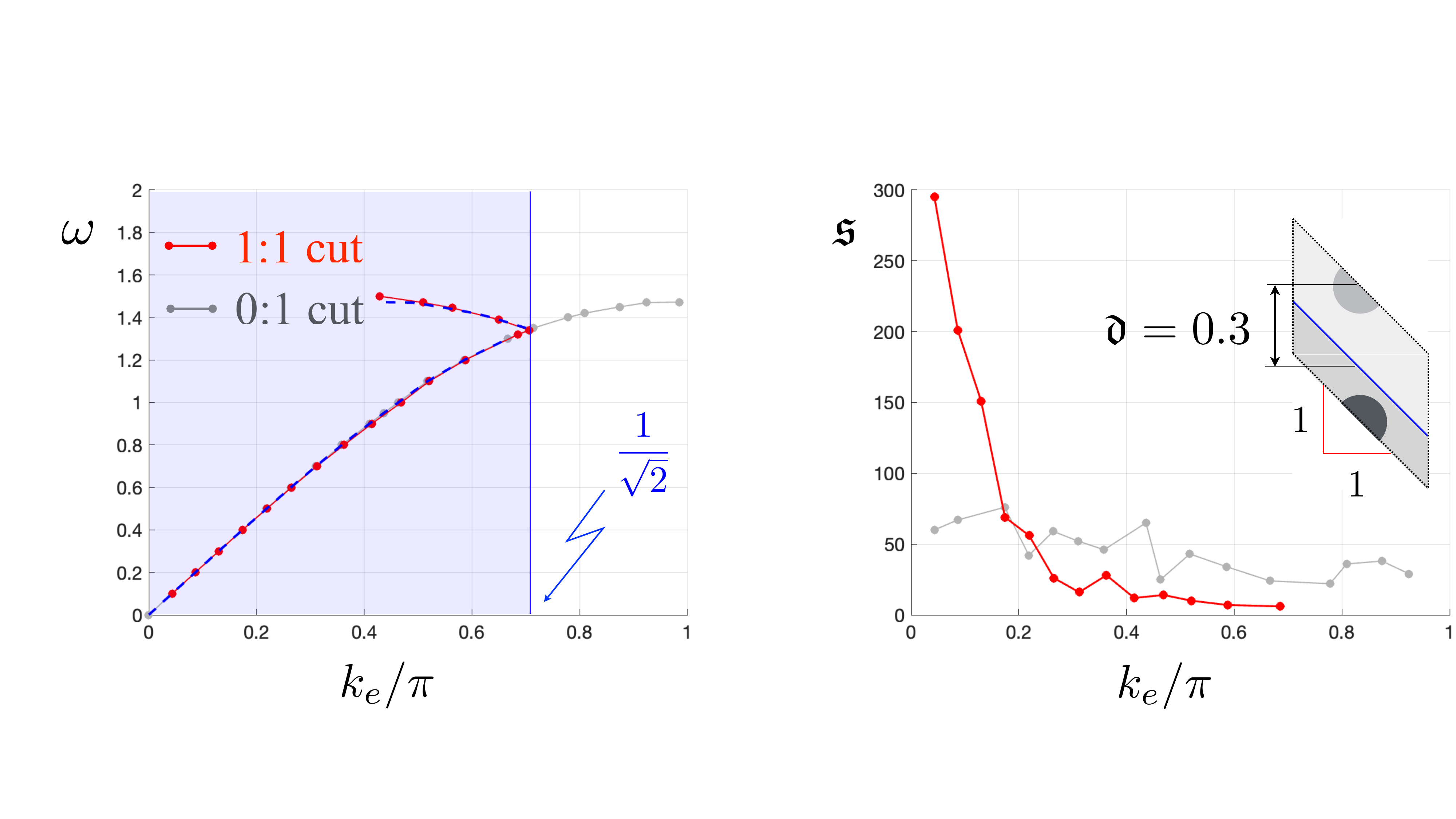}}
\caption{Propagation of SB waves along the 1:1 rational surface cut made at depth $\mathfrak{d}=0.3$: dispersion diagram (left panel), and variation of skin depth vs.~$k_e$ due to~\eqref{skindepth1a}--\eqref{skindepth1b} (right panel). As a point of reference, results for the 0:1 horizontal cut at depth $\mathfrak{d}=0.3$ are plotted alongside in light gray. In the left panel, shaded region indicates the positive half of the first Brillouin zone for the 1:1 cut, while the dashed blue lines plot the 0:1 dispersion diagram folded onto the support of the 1:1 Brillouin zone, $(0,1/\sqrt{2})$. The inset shows the  multi-cell~$\tilde{Y}$ corresponding to the 1:1 cut.} \label{non-orth-cut}
\end{figure}

Our last example illustrates the boundary layer due to the free surface cut~$\mathcal{S}$ made at depth \mbox{$\mathfrak{d}=0.3$} with 1:1 slope ($q^1 = q^2 = 1$).  The resulting unit cell $\tilde{Y}$, shown as inset in Fig.~\ref{non-orth-cut}, is used in~\eqref{QEP-surf3} to compute the germane QEP eigenspectrum. The left panel in Fig.~\ref{non-orth-cut} shows in red bullets the first and second branch of the SB dispersion diagram for the 1:1 cut \mbox{($0<k_e/\pi<1/\sqrt{2}$)}, contrasted against the 0:1 horizontal cut with $\mathfrak{d}=0.3$ shown in gray ($0<k_e/\pi<1$). Due to extended support of the multi-cell $\tilde{Y}$, the first Brillouin zone for the 1:1 cut is $1/\sqrt{2}$ times that for the 0:1 cut. For the ease of comparison, blue dashed lines plot the 0:1 dispersion diagram \emph{folded} onto the support of the 1:1 Brillouin zone. As can be seen from the display, the first two branches of the 1:1 dispersion diagram nearly coincide with the folded first branch of the 0:1 dispersion diagram. Here it is useful to observe that every eigenpair ($\kappa_n,\phi_n$) of the QEP~\eqref{QEP-surf2} computed for the unit cell~$Y$ (0:1 slope) generates -- by the periodic extension of $\phi_n$ -- an eigenpair ($\tilde{\kappa}_m=\kappa_n,\tilde{\phi}_m$) of the QEP~\eqref{QEP-surf3} for the multi-cell~$\tilde{Y}$ (1:1 slope). Notwithstanding this property, the near-overlap of the two SB dispersion diagrams in Fig.~\ref{non-orth-cut} is a peculiarity of the featured configuration and is not expected to hold in general. For completeness, the right panel in Fig.~\ref{non-orth-cut} plots the skin depth variation (due to quantal estimate~\eqref{skindepth1a}--\eqref{skindepth1b}) of the SB waves for the 1:1 and 0:1 slopes, demonstrating a marked effect of cut orientation on character  of the boundary layer.

\section{Summary}

\noindent In this study, we establish a basic framework for the analysis of boundary layers, termed surface Bloch (SB) waves, in periodic media truncated along a plane with rational slope relative to the ``mother'' Bravais lattice. To provide a focus for the analysis, we consider a 2D scalar wave equation with periodic coefficients (describing antiplane shear waves in phononic crystals) and assume homogeneous Neumann data on the cut surface. Our analysis revolves around a quadratic eigenvalue problem (QEP) for the effective unit cell of periodicity -- deriving from a geometric interplay between the mother Bravais lattice and (rational) orientation of the surface cut -- that seeks the complex wavenumber controlling the evanescence away from the cut plane given (i) the excitation frequency and (ii) wavenumber in the direction of the surface cut. In this way, the boundary layer is obtained as a linear combination of the evanescent QEP eigenstates, whose relative amplitudes are computed by imposing the boundary condition on the cut surface. With the QEP eigenspectrum at hand, evaluation of an SB wave -- in terms of both dispersion characteristics and evanescent waveforms -- entails only a low-dimensional eigenvalue problem. This feature caters for rapid exploration of the effect of (periodic) surface undulations, and so enables manipulation of the SB waves via optimal design of the surface cut. The proposed framework is readily extendable to other types of boundary conditions, governing equations (e.g. Navier, Maxwell), and interfacial waves arising between bonded periodic media. 

\bibliographystyle{plain}
\bibliography{refs} 

\end{document}